\documentclass[aps,pra,twocolumn,superscriptaddress,longbibliography,nofootinbib]{revtex4-2}
\usepackage{amsmath, amsfonts, amsthm, amssymb}
\usepackage[margin=1in]{geometry}
\usepackage{graphicx,xcolor,tikz}
\usepackage[hidelinks]{hyperref}
\usepackage[caption=false]{subfig}

\usepackage{qcircuit}

\usepackage{dcolumn}
\usepackage{blindtext}
\usepackage{floatrow}
\usepackage[version=3]{mhchem}
\usepackage{footnote}
\usepackage{bm}
\usepackage{dsfont}
\usepackage{xr-hyper}
\usepackage{cleveref}
\usepackage{soul}
\usepackage[normalem]{ulem}

\usepackage{hhline}
\usepackage{booktabs}
\usepackage{algorithm}
\usepackage[noend]{algpseudocode}

\Crefname{equation}{Eq.}{Eqs.}
\Crefname{equation}{Equation}{Equations}
\Crefname{figure}{Fig.}{Figs.} 
\Crefname{figure}{Figure}{Figures}
\Crefname{section}{Sect.}{Sects.}
\Crefname{section}{Section}{Sections}
\Crefname{table}{Table}{Tables}
\Crefname{appsec}{}{Appendices}

\labelformat{subsection}{\thesection.#1}
\labelformat{subsubsection}{\thesection.\thesubsection.#1}

\begin{document}

\title{Snakes and Ladders: Adapting the surface code to defects}

\author{Catherine Leroux}
\email{catilero@amazon.com}
\author{Sophia F. Lin}
\author{Przemyslaw Bienias}
\author{Krishanu R. Sankar}
\affiliation{AWS Center for Quantum Computing, Pasadena, CA 91125, USA}
\author{Asmae Benhemou}
\affiliation{AWS Center for Quantum Computing, Pasadena, CA 91125, USA}
\affiliation{Department of Physics and Astronomy, University College London, London WC1E 6BT, United Kingdom}
\author{Aleksander Kubica}
\affiliation{AWS Center for Quantum Computing, Pasadena, CA 91125, USA}
\affiliation{Yale Quantum Institute \& Department of Applied Physics, Yale University, New Haven, CT 06520, USA}
\author{Joseph K. Iverson}
\affiliation{AWS Center for Quantum Computing, Pasadena, CA 91125, USA}

\begin{abstract}
    One of the critical challenges solid-state quantum processors face is the presence of fabrication imperfections and two-level systems, which render certain qubits and gates either inoperable or much noisier than tolerable by quantum error correction protocols.
    To address this challenge, we develop a suite of novel and highly performant methods for adapting surface code patches in the presence of defective qubits and gates, which we call \emph{Snakes and Ladders}. We explain how our algorithm generates and compares several strategies in order to find the optimal one for any given configuration of defective components, as well as introduce heuristics to improve runtime and minimize computing resources required by our algorithm. In addition to memory storage we also show how to apply our methods to lattice surgery protocols. Compared to prior works, our methods significantly improve the code distance of the adapted surface code patches for realistic defect rates, resulting in a logical performance similar to that of the defect-free patches.
\end{abstract}

\maketitle

\begin{figure*}[ht!]
    \centering
    \includegraphics[width=.93\linewidth]{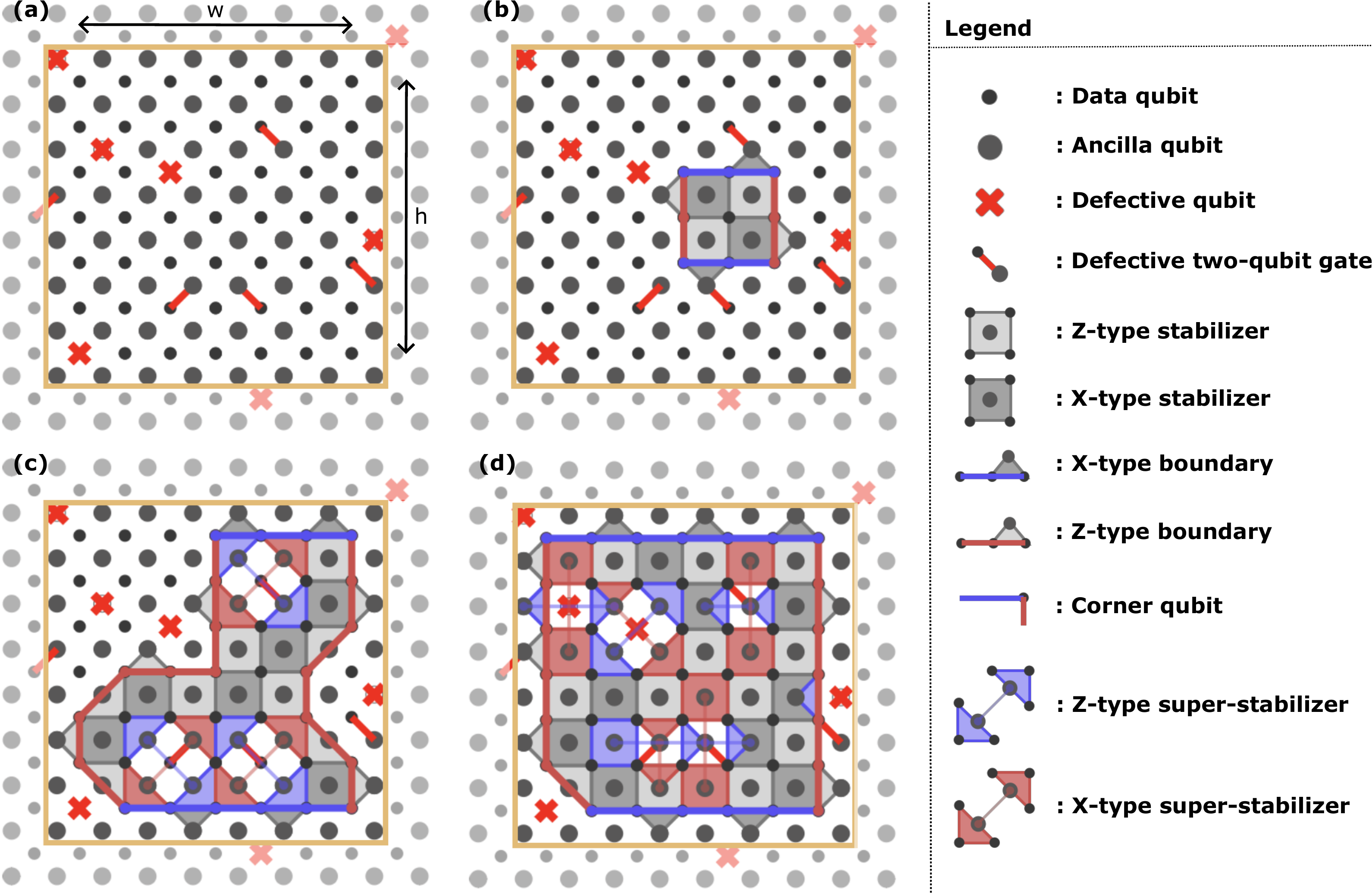}
    \caption{(a) A processor consisting of a square lattice of qubits with nearest-neighbor gates. Several defective components marked in red. The question we study is how to realize the most performant surface code within a target window (yellow square) of dimensions $w\times h$. In absence of defects, this window supports a surface code with distance up to $(d_X, d_Z) = (h, w)$. In this example $w=h=7$. (b) A surface code patch with distance $(d_X, d_Z) = (3, 3)$ featuring no defective components. This is the surface code patch with the largest distance, free of defective components, that we can make in the current window. (c) A surface code patch adapted to defects using DQD. This code has distance $(d_X, d_Z) = (5, 3)$. (d) A surface code patch adapted to defects using SnL, resulting in distance $(d_X, d_Z) = (6, 6)$ which is closer to the maximum distance possible in the target window. In (b)-(d), gauge checks of the same color are combined together into super-stabilizers and are connected together with lines.}
    \label{fig:sprinked_defects}
\end{figure*}

\section{Introduction}

    One of the leading architectures for error-corrected quantum computation is based on the surface code~\cite{Kitaev03,Bravyi1998,Dennis02,fowler2012surface} and can be realized with qubits on a square grid and nearest-neighbor entangling gates.
    This approach is particularly attractive for superconducting qubits, which are typically arranged on a two-dimensional plane and strongly coupled to their nearest neighbors. However, hardware implementation of the surface code is susceptible to defective qubits or gates, which can arise due to mistargeted frequencies, fabrication defects~\cite{kjaergaard2020superconducting,bilmes2020resolving}, or two-level systems (TLSs)~\cite{klimov2018fluctuations,burnett2014evidence,muller2019towards}.
    Effectively adapting the surface code to accommodate these imperfections has emerged as a critical challenge in the effort to scale up superconducting qubit devices.
    
    Prior work~\cite{auger2017fault,strikis2023quantum,siegel2023adaptive,lin2024codesign} has focused on one particular strategy of handling defective qubits and gates, which we collectively refer to as data-qubit disabling (DQD). On top of defective data qubits, DQD also disables data qubits that are either involved in defective two-qubit gates or next to defective ancilla qubits. DQD converts the stabilizer checks around defective or disabled data qubits to gauge checks, which then comprise super-stabilizers. Recent refinements of DQD~\cite{yin2024flexiscd,wei2024low} optimized super-stabilizers to minimize reduction of the surface code distance for certain configurations of defects. Nonetheless, reduction of the distance remains significant and therefore appreciably diminishes the logical performance of the adapted surface code. 
    
    In this article, we introduce a new framework, called ``Snakes and Ladders'' (SnL), to adapt the surface code in the presence of defective qubits and gates. Compared to prior works, SnL greatly improves the code distance of the adapted surface code while ensuring that all stabilizers and super-stabilizers are measured over two error-correction rounds. Consequently, SnL reduces the qubit overhead of memory storage and fault-tolerant computation via lattice surgery with the surface code~\cite{horsman2012,Litinski2018latticesurgery} in the presence of defective components. 
    
    SnL makes use of a novel primitive operation for defective ancilla qubits and two-qubit gates, called ``ancilla repurposing''. Ancilla repurposing uses two of the neighboring ancilla qubits to measure weight-two gauge checks that, together, allow to recover the value of the missing stabilizer check. This primitive generally preserves the code distance, offering a significant gain over DQD (which reduces the distance by two for a defective ancilla qubit and by one for a defective two-qubit gate).
    
    By ingeniously composing the primitives operations, SnL constructs many global strategies for any configuration of defective components, resulting in the adapted surface code with the optimal code distance. SnL not only handles the defective components in the bulk; it also optimally deforms the boundary of the surface code. To reduce the computational overhead of constructing and analyzing global strategies, SnL may use heuristic methods to reduce the search space.

    We call our algorithm ``Snakes and Ladders'' because repurposing ancillas (which helps us preserve code distance) is analogous to climbing ladders whereas disabling data qubits (which reduces the code distance) is akin to sliding down snakes in the eponymous game~\cite{board_game}. As explained in \Cref{sec:neighboring_defects}, several repurposed checks in a row resemble the rungs of a ladder, and the strings of disabled data qubits resemble snakes.
    
    The article is organized as follows. First, we provide motivation in~\Cref{sec:motivation} and describe the key steps of SnL in~\Cref{sec:algorithm}. Next, in~\Cref{sec:Results} we report numerical simulations for the surface code adapted to random configurations of defective components. In particular, we analyze memory storage, provide yield estimates and benchmark a logical CNOT gate implemented via lattice surgery. We conclude by discussing the impact of our work and some future directions in~\Cref{sec:conclusion}.

\section{Motivation}\label{sec:motivation}

    Suppose that we have a processor consisting of a square lattice of data and ancilla qubits with nearest-neighbor gates. We wish to realize one or more patches of surface code within certain regions of the processor to enable quantum information processing, e.g. with lattice surgery. We restrict ourselves to the rotated surface code, which realizes the same code distance as the unrotated surface code with fewer physical qubits. We choose an orientation with $X$-type boundaries on top and bottom and $Z$-type boundaries on left and right. The vertical (horizontal) distance of the code is denoted $d_X$($d_Z$).
    
    In what follows, we call a rectangular region of our processor in which we wish to realize a surface code patch a ``target window." A target window is said to have dimensions $w \times h$ with width $w$ and height $h$ if it contains a grid of $w \cdot h$ data qubits and $(w+1)\cdot (h+1)$ ancilla qubits such that only ancilla qubits are on the perimeter of the window. \Cref{fig:sprinked_defects}(a) shows a target window (yellow square). In the absence of defective components, this window supports a rotated surface code with distance up to $(d_X,d_Z) = (h, w)$. We use the terms ``defect" or ``defective component" to refer to a qubit or two-qubit gate in our surface code lattice that we cannot use in our code, either because it is non-operational or because it is so much noisier than the other qubits or two-qubit gates that we prefer not to use it. We use ``data defect", ``ancilla defect", and ``link defect" to refer to defective data qubits, ancilla qubits, and two-qubit gates, respectively.

    Suppose that our processor has defective components and that we have characterized the defects, indicated in~\Cref{fig:sprinked_defects}(a) with red marks on qubits and links, within the target window (yellow square). Then, we must determine how to construct a surface code patch within this target window. One approach would be to search for the surface code patch with the highest distance that is not compromised by defects within the target window. In general, defects will be distributed across the entire target window, and this approach will produce a small patch, like in \Cref{fig:sprinked_defects}(b). In this example, the resulting surface code patch has distance $(d_X, d_Z) = (3, 3)$ in a $7\times 7$ target window. More generally, we would be unable to implement lattice surgery operations between this patch and patches in other windows unless we can find a defect-free routing space between them. To get larger code distance and to enable lattice surgery, we require a strategy for adapting our surface code patches to the presence of defects.

    We place three constraints on our defect strategies. First, we insist on preserving the number of encoded qubits and the topology of the boundaries. Second, we require that the full set of stabilizer generators can be measured in two error-correction rounds, just like in prior approaches \cite{auger2017fault,strikis2023quantum,siegel2023adaptive,lin2024codesign,yin2024flexiscd,wei2024low}. Finally, we do not use weight-1 gauge checks~ \cite{yin2024flexiscd,wei2024low}. We find that incorporating weight-1 checks in SnL often leads to more noise and no improvement in code distance (see \Cref{app:weight1_checks} for more detail). Subject to these constraints, we aim to maximize the code distance of the adapted surface code.

    Two such strategies are shown in \Cref{fig:sprinked_defects}(c) and (d). Panel (c) is the strategy we find using DQD. This surface code patch has distance $(d_X, d_Z) = (5,3)$, improving slightly over the code patch in panel (b). Panel (d) shows the optimal strategy we find with SnL. Here the patch has distance $(d_X, d_Z) = (6, 6)$. This patch is expected to result in smaller logical error rates than the patch generated with DQD since the performance is primarily limited by the minimum of the $X$- and $Z$-distance of the code. Equivalently, if we require a certain minimum logical performance for our patch, SnL allows us to work with a smaller target window than DQD. In turn, this enables more efficient layout of logical qubits in our processor, reducing the overhead of fault-tolerant quantum computation.

    \begin{figure}[ht!]
        \centering
        \includegraphics[width=\linewidth]{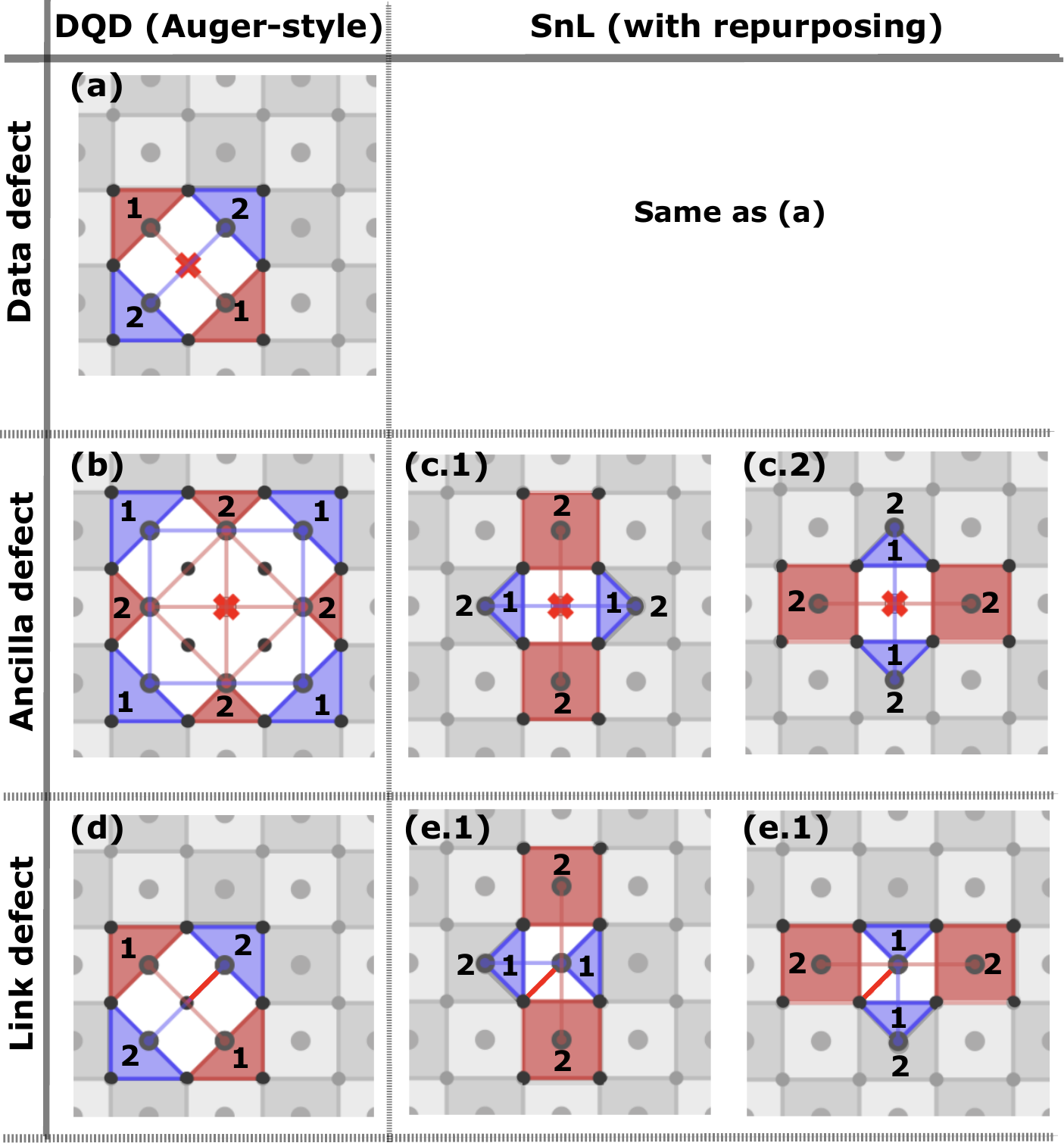}
        \caption{Primitive operations used in our framework, for isolated defects. The highlighted checks are measured in alternating rounds, and the numbers indicate which checks are measured in odd and even rounds. The DQD strategies are shown in the left column for (a) a data defect, (b) an ancilla defect, (d) a link defect (mapped to a data defect). The SnL strategies for the same defects are shown in the right column.}
        \label{fig:primitive_operations}
    \end{figure}

\section{Snakes and Ladders (SnL) framework}\label{sec:algorithm}

    In this section, we detail the key steps of our framework for maximizing surface code distance in the presence of defects, which is fully implemented and available in our repository~\cite{SnL2024}.
    
    \subsection{Strategies for isolated defects}\label{subsection:isolated_defects}
    
        For a given configuration of defects, our framework constructs many possible strategies to adapt the surface code, and selects the one with the best code distance. In this section, we present our strategies supposing we have only a single defect in our target window (see \Cref{fig:primitive_operations}). For conciseness, we focus on defective components in the bulk, but the approach can be straightforwardly generalized to defects on the boundary (see~\Cref{app:efficient_check_boundary} for details). Our strategies are designed such that $X$-type and $Z$-type super-stabilizers are measured every other round. The unmodified stabilizer checks of the surface code are still measured every round.

        \subsubsection{Data defect}
    
            The primitive for isolated data defects is illustrated in~\Cref{fig:primitive_operations}(a). The prescription is to remove the defective data qubit from the four adjacent stabilizer checks, converting them to four weight-3 checks. After this change, neighboring checks of different Pauli type anti-commute, and therefore can no longer be measured simultaneously. This necessitates the use of an alternating schedule where the $X$-type checks and $Z$-types checks are measured in separate rounds. The measurement outcomes from individual gauge checks are random, but the product of the two $X$-type ($Z$-type) checks is deterministic and forms a single $X$-type ($Z$-type) stabilizer of weight 6. This approach ``patches" the hole in the code and preserves the number of encoded logical qubits. However, the distance of the code is reduced by 1 in both directions because the lengths of $X$ and $Z$ logical strings passing through the hole in the code are both reduced by 1. Here the term ``hole'' denotes a set of inter-dependent super-stabilizers whose gauge checks commute with the checks outside the set but not with all the gauge checks within the set.
        
        \subsubsection{Ancilla defect}\label{sec:ancilla_repurposing}
    
            Ancilla repurposing is a new primitive we introduce to handle isolated defective ancilla qubits. Two possible ways of using ancilla repurposing for an isolated ancilla defect are shown in~\Cref{fig:primitive_operations}(c). Here, the weight-4 stabilizer of the defective ancilla is split into two weight-2 gauge checks, and neighboring ancilla qubits on the left and right (top and bottom) are “repurposed” to measure these gauge checks in alternate rounds (in addition to their own stabilizer checks). The two weight-4 checks directly above and below (left and right) anti-commute with these weight-2 gauge checks and are thus themselves gauge checks which multiply to form one weight-8 stabilizer. Both left/right and top/bottom ways of repurposing around an ancilla defect are valid. However, depending on the Pauli type of the check the defective ancilla is supposed to measure, one orientation preserves the code distance, whereas the other reduces it by two along the weight-8 stabilizer (see~\Cref{app:efficient_check_distance}). We refer to one orientation as ``distance-preserving" and the other as ``lossy." In contrast, DQD disables all four data qubits in the neighborhood of the ancilla defect (see~\Cref{fig:primitive_operations}(b)), reducing $d_X$ and $d_Z$ by 2.
            
            We find that ancilla repurposing benefits from what we call ancilla padding. Instead of using a grid with physical qubits at every other lattice position around the boundary (locations of the weight-2 checks in the defect-free surface code), we work with a target window that has a physical qubit at every position around the boundary as in \Cref{fig:sprinked_defects}. This allows us to repurpose these ancillas to recover the full code distance for ancilla defects near the boundary that we otherwise could not handle without loss of distance. See~\Cref{app:efficient_check_boundary} for an illustration.

        \subsubsection{Link defect}
        
            In the case of an isolated defective two-qubit gate, the ancilla qubit involved in the gate can still measure a weight-2 check that is part of its native stabilizer while one of the neighboring ancilla qubits can measure the other weight-2 check needed for recovering the original weight-4 stabilizer. This is illustrated in~\Cref{fig:primitive_operations}(e). As with ancilla defects, we have two orientations for the repurposing, one distance-preserving and one lossy. Similarly to ancilla defects we can recover the code distance, whereas DQD disables the data qubit in the defective gate (see~\Cref{fig:primitive_operations}(d)), leading to a distance loss of 1 in both directions.
    
    \subsection{Strategies for clustered defects} \label{sec:neighboring_defects}
    
        \begin{figure*}[ht!]
            \centering
            \includegraphics[width=\linewidth]{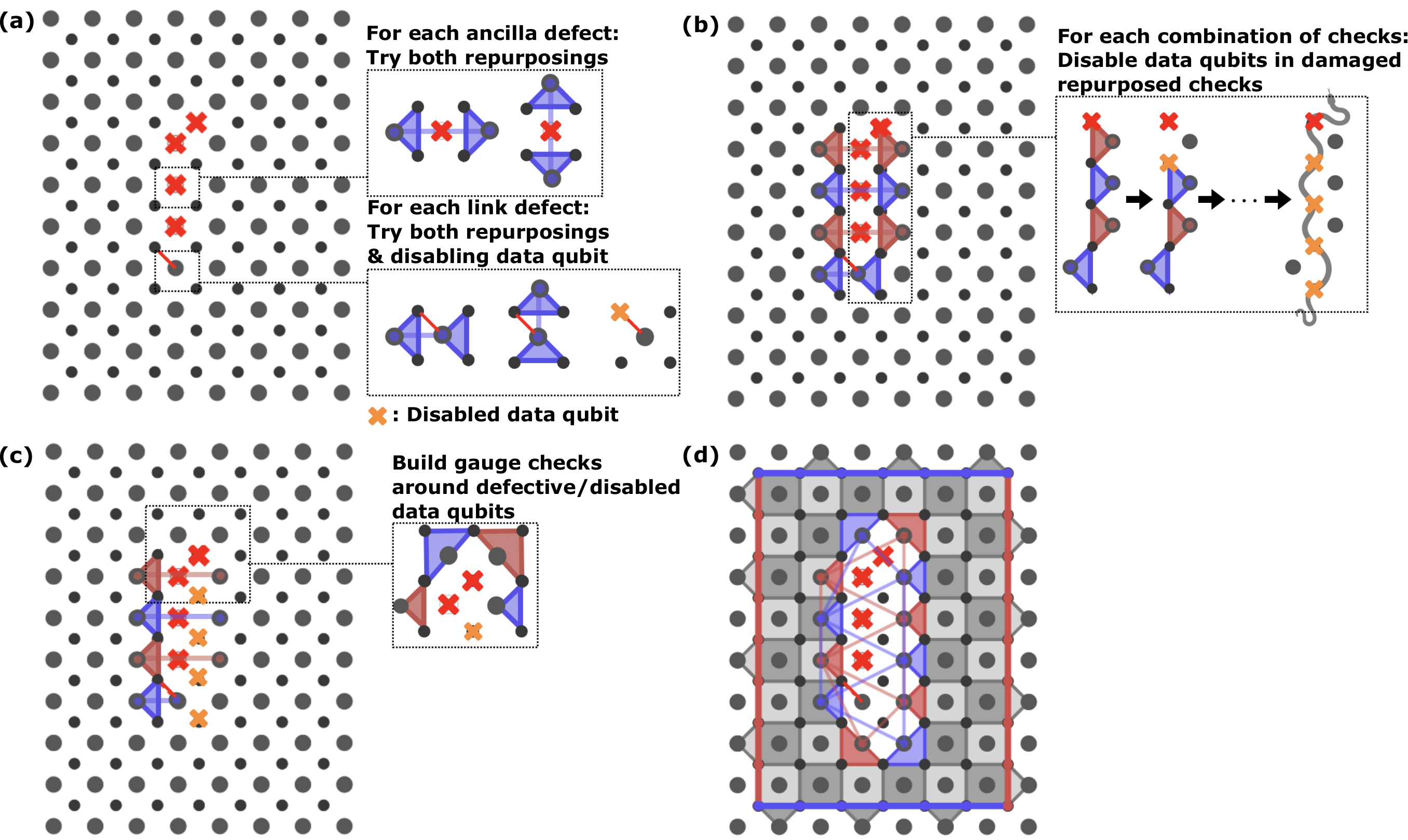}
            \caption{Applying SnL to a cluster of defects. (a) We identify ancilla and link defects and combine all possible ways of repurposing.(b) For each combination of primitive operations found in (a) we determine which repurposed checks are invalid and disable data qubits in them. This process is simplified by identifying connected components of repurposed checks. We disable all data qubits in a connected component if there is a defect anywhere in that component or if an ancilla qubit is repurposed more than once. These disabled data qubits form a ``snake''. (c) We apply the data defect primitive around each defective and disabled data qubit, converting stabilizers around them to gauge checks. (d) We identify the super-stabilizers based on the commutation relations. For more details, see \Cref{app:super-stabilizers}.}
            \label{fig:snl_snakes}
        \end{figure*}
        
        Naively, we would apply the distance-preserving repurposing for every ancilla and link defect in the patch as if they were isolated. However, this is not always possible when multiple defects are close together. Neighboring defects that cannot be considered separately using the strategies in \Cref{fig:primitive_operations} must be grouped into ``defect clusters'' and considered collectively. Within a defect cluster, repurposing a given ancilla might not be possible for two reasons. First, we do not allow an ancilla qubit to be repurposed more than once, since doing so would make it impossible to measure all stabilizers over two measurement rounds. Second, we do not allow repurposed weight-1 checks (see \Cref{app:weight1_checks} for a detailed explanation), implying that any defective component (data, ancilla or gate) in a repurposed weight-2 check makes this check invalid. To solve this issue, we disable all data qubits involved in tentative repurposed checks that would be compromised by defects. This process results in SnL strategies that mix ancilla repurposing and the data defect primitive in~\Cref{fig:primitive_operations}(a).
        
        Therefore, in an attempt to minimize the number of disabled data qubits, our framework ultimately considers both orientations of repurposing for each ancilla (\Cref{fig:primitive_operations}(c)) and link defect (\Cref{fig:primitive_operations}(e)); in the presence of other defects, we might choose the lossy repurposing to optimize the overall code distance. We remark that if the ancilla in a link defect cannot be used, either because it is itself defective or due to another defective gate, we revert back to the same strategies as for an ancilla defect. Moreover, for every link defect, we also consider the possibility of disabling the data qubit instead of repurposing (\Cref{fig:primitive_operations}(d)). This can sometimes result in fewer data qubits being disabled, for example, when the data qubit in a defective gate is also defective.
    
        Defect clusters grow quickly in number and in size with respect to the defect rate~\cite{wei2024low}. Consequently, we need a way to maximize the number of repurposed checks, disabling a minimum of data qubits. \Cref{fig:snl_snakes} illustrates how SnL finds valid strategies by combining both primitives for any defect configuration. The first step is to combine all possible strategies for the isolated ancilla and link defects as shown in panel (a) for an example defective configuration. Given a specific combination of repurposed weight-2 checks, we have a systematic way to determine which repurposed checks are compromised and which data qubits to disable. We define a graph $\mathcal{G}$ whose vertices are all the qubits (both ancilla and data) involved in these weight-2 checks. Each edge in $\mathcal{G}$ is a two-qubit gate in one of those checks. Any defect (data, ancilla or gate) or ancilla repurposed more than once in a connected component of $\mathcal{G}$ results in all the data qubits in that same component to be disabled. This ensures that we have no weight-1 check left after repurposing and no ancilla repurposed more than once. We call those strings of disabled data qubits ``snakes''. 
        
        This process is illustrated in panel (b) where a single data defect results in all data qubits in a chain of repurposed checks to be disabled. After having defined which repurposed checks are feasible, we then apply the data defect primitive around each defective and disabled data qubit, i.e., we convert each damaged stabilizer around them to gauge checks. This step is shown in panel (c). We then define the super-stabilizers by combining the remaining repurposed checks with the gauge checks around the disabled data qubits. One of the many possible strategies for the example defect configuration in (a) is shown in (d).
    
        \begin{figure}[ht!]
            \centering
            \includegraphics[width=\linewidth]{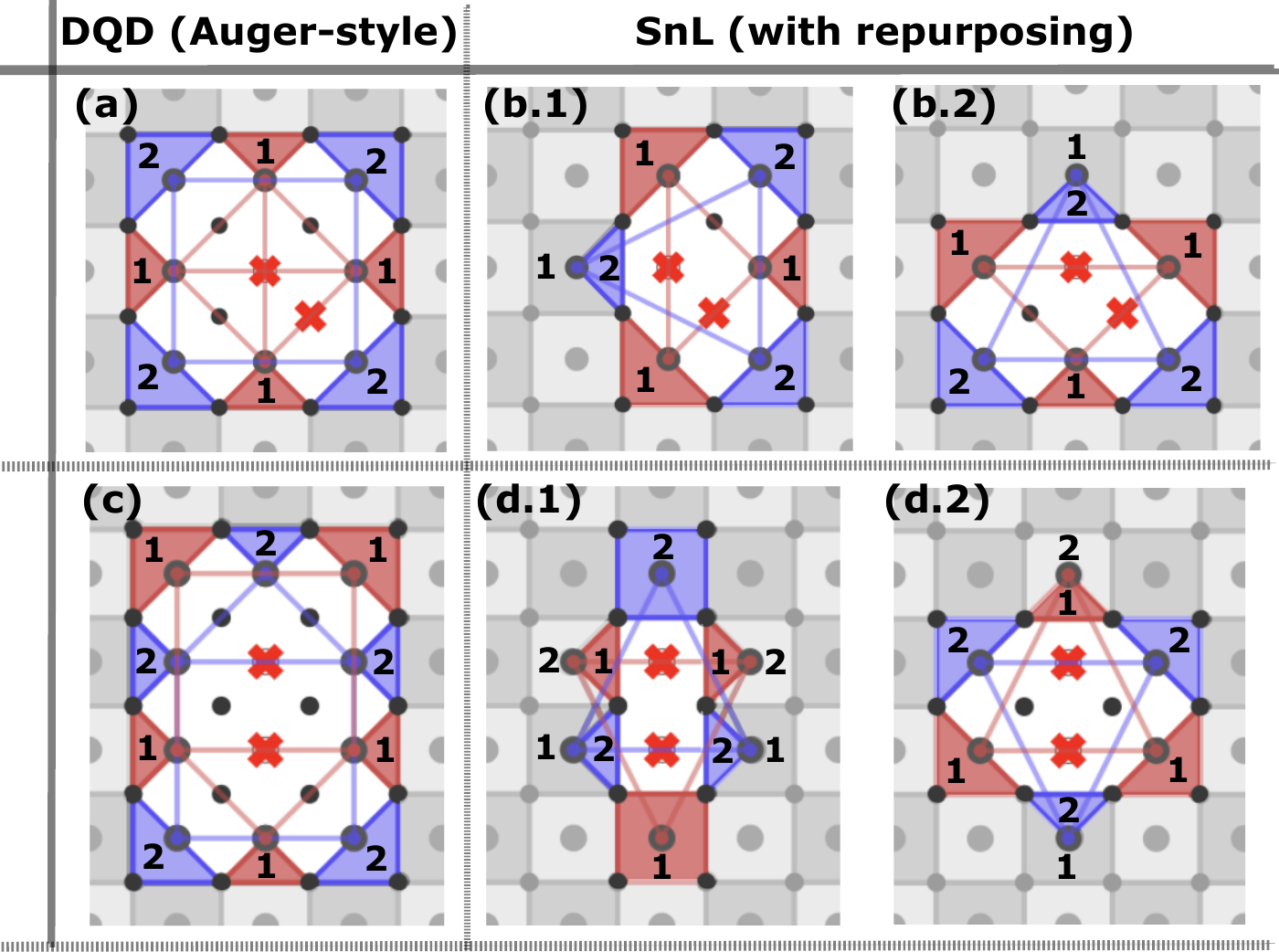}
            \caption{Application of our strategies for two examples of defect clusters: (a)-(b) adjacent ancilla and data defects, and (c)-(d) two adjacent ancilla defects. (b) and (d) show the best two strategies that we find for each defect configuration with SnL.}
            \label{fig:two_examples}
        \end{figure}
        
        Two additional examples are shown in~\Cref{fig:two_examples}. Panel (b) shows two of the possible strategies SnL returns for adjacent ancilla and data defects. The data defect renders one of the two repurposed checks invalid: we are therefore left with a single repurposed check and a disabled data qubit. These two strategies still result in a better code distance than DQD in panel (a), where all four data qubits are disabled. Another example is shown in~\Cref{fig:two_examples}(d). There, in order to avoid repurposing twice, we need to either consider a lossy repurposing (opposite to the orientation that preserves code distance) for one of the two ancilla qubits in panel (d.1) or disable two data qubits in panel (d.2). DQD, on the other hand, leads to 6 data qubits being disabled.
    
    \subsection{Defect clusters on the boundary}\label{subsection:boundary_deformation}
    
        \begin{figure}[ht!]
            \centering
            \includegraphics[width=\linewidth]{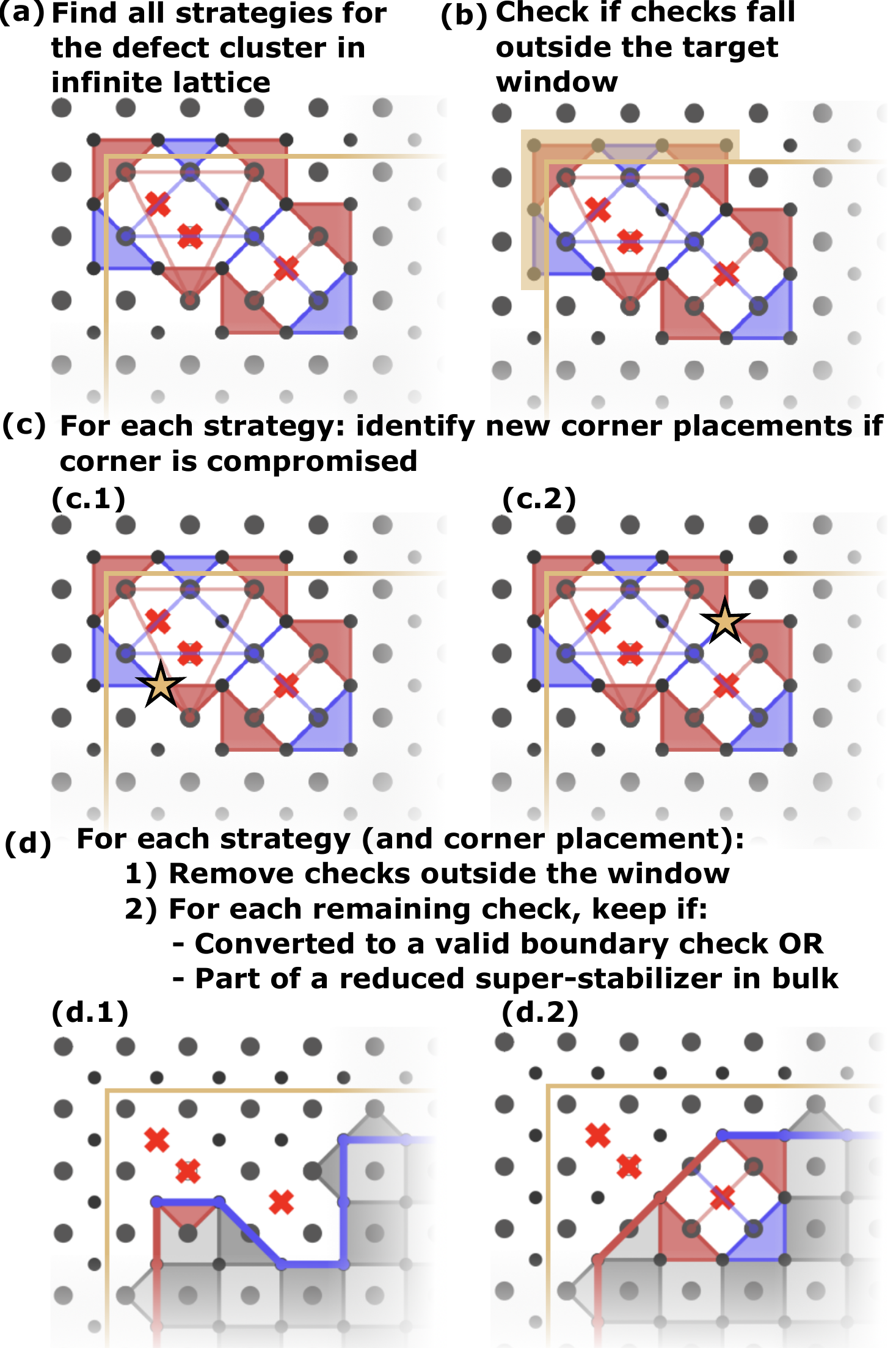}
            \caption{Routine for the boundary deformation. (a) We find all strategies for the defect cluster. We only show the top left corner of the target window  (in yellow) for compactness. b) For each strategy, we determine if the gauge checks require qubits outside of the target window (highlighted in yellow). (c) For each strategy, we determine if one or more corners are part of the gauge checks or disabled. Here we show (using the yellow star) two possible corner placements for the strategy in (b). (d) If a corner is compromised, we repeat this step for every possible corner placement found in (c), otherwise we only do this step once. We remove all gauge checks that fall outside the target window. For each remaining gauge check we determine if it can 1) be converted to a valid boundary check or 2) be part of a smaller super-stabilizer within the target window such that the commutation relations are satisfied. If not, then we eliminate the check. For example, the two red gauge checks in (c.1) became gray boundary checks in (d.1). On the other hand, the red gauge checks in (d.2) form a new super-stabilizer in the bulk.
            }
            \label{fig:boundary_deformation}
        \end{figure}
    
        Up to this point, we have ignored the impact of the target window's boundary on SnL. We now discuss how we deform the boundary of the code patch when the defect cluster touches the boundary of the target window, in a way that maximizes the code distance.
    
        SnL first entails embedding the original target window in an infinite grid we call the ``extended window", such that there are no defects outside the target window. This simplifies SnL by allowing it to treat bulk and boundary defects equally. \Cref{fig:boundary_deformation}(a) shows an example where we have defined a target window that contains a cluster of two data defects and one ancilla defect. Moreover, one of the data defects is a corner qubit, which we define as a data qubit where an $X$-type boundary meets with a $Z$-type boundary. In this example, one possible strategy for this defect cluster includes some checks outside the target window. These checks are invalid and need to be removed. The remaining checks around the cluster must be converted to valid boundary checks, unless they become part of a reduced super-stabilizer in the bulk.
        
        Before adapting the checks in the example defect cluster, we need to move the corner from the defective data qubit to a functional one. The position of the new corner determines what type of boundary check is valid, at each point of the new boundary. The next step of this routine is therefore to identify all new possible corner positions for each strategy we found for the cluster. This step is illustrated in panel (c). To the best of our knowledge, the optimization of the corner placement has not been considered before, but it helps both SnL and DQD achieve better code distance. 
        
        For each of these new corner positions, we must determine which checks need to be removed from the strategy and which remaining checks need to be combined into smaller super-stabilizers. This step is depicted in panel (d). We first remove all checks (repurposed and gauge) that fall outside the target window. For each remaining check we must determine whether it can be 1) converted to a valid boundary check (a check of the Pauli type corresponding to that of the boundary) or 2) combined with other checks to form a reduced super-stabilizer inside the target window, such that the stabilizers all commute. If none of these two options is possible, we remove the check. In~\Cref{app:boundary_deformation} we provide a systematic way that maximizes the number of data qubits inside the target window.
    
        We remark that the extended window need not be an infinite lattice, but rather it is enough if it contains two additional layers of data qubits and ancilla qubits on each side of the target window. That way, the data defect primitive in \Cref{fig:primitive_operations}(a) is supported for an ancilla defect on the boundary of the target window. It follows that all other primitives in \Cref{fig:primitive_operations} are supported for any possible defect in the target window.
        
        Finally, we emphasize that our implementation for the boundary deformation minimizes the number of data qubits that are removed from the code for each strategy. However, in certain cases it is possible to construct a larger-distance code in a smaller area of the target window, for example, by excluding a region with a high density of defects. However, this defers the problem of adapting to all the defects in the target window until we attempt to do lattice surgery with patches outside the target window.
        
    \subsection{Recap of SnL on a defective patch}
    
        \begin{figure}[ht!]
            \centering
            \includegraphics[width=\linewidth]{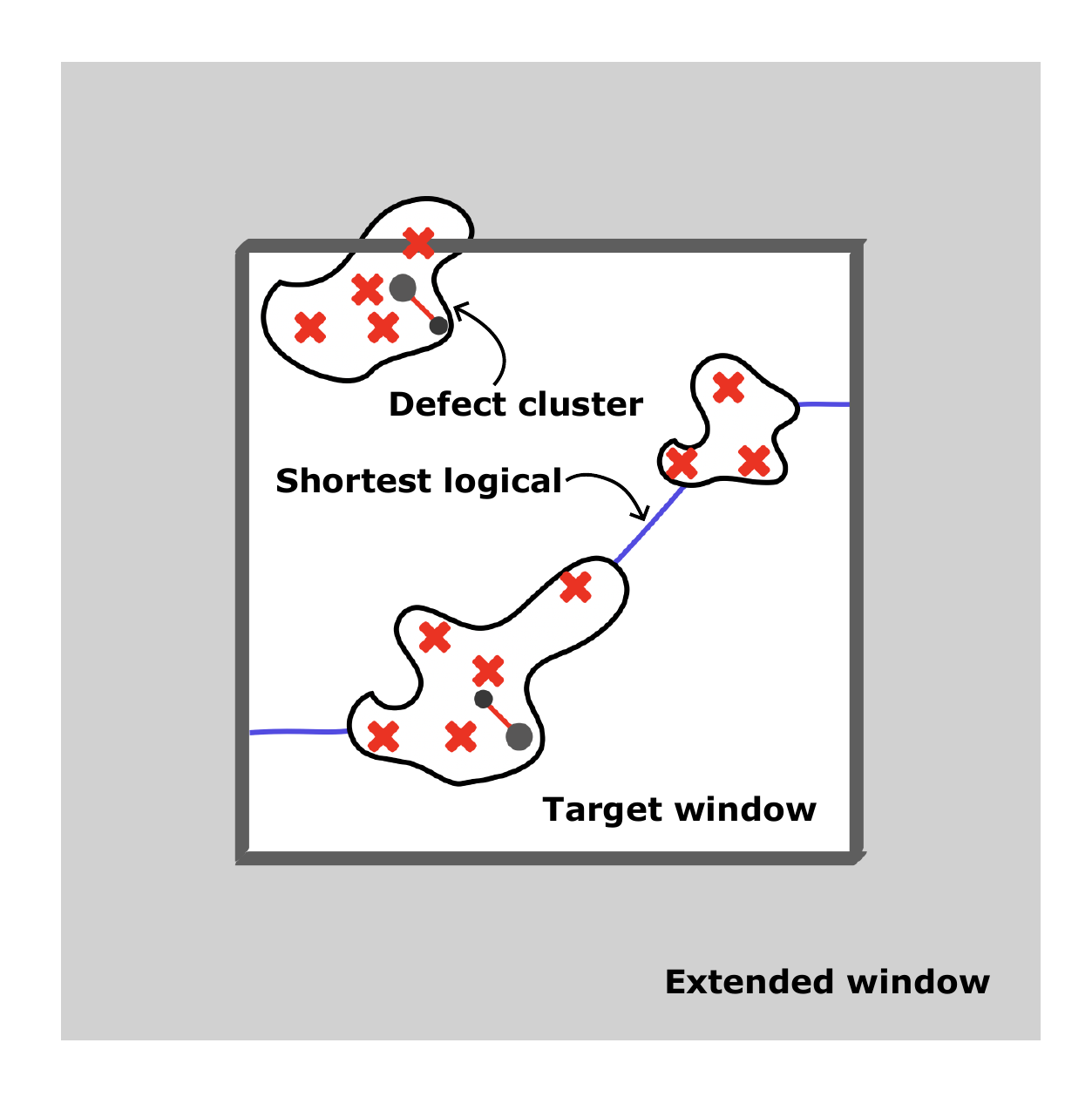}
            \caption{Cartoon of a processor with multiple defect clusters. For the purpose of the algorithm, we embed a target window in an extended window, which is simply a trick to facilitate our routine for clusters near the boundary. We define defect clusters in the extended window and find all strategies for each cluster individually. We then combine these strategies to form global strategies that maximize the code distance that can be realized within the target window.}
            \label{fig:defective_grid}
        \end{figure}
    
        In this section, we describe how we find the best global strategies for a target window of a defective processor (see \Cref{fig:defective_grid}). Here, we imagine having a defective processor where a random defect configuration can be decomposed into several defect clusters. The goal here is to find a global strategy that maximizes the distance of the surface code patch.
    
        In \Cref{subsection:isolated_defects,sec:neighboring_defects}, we explained how to find local strategies for an isolated defect or a defect cluster. For each local strategy, we can compute the ``local distance loss," which is the distance of the surface code patch adapted to this single defect or defect cluster, relative to the maximum distance achievable in the defect-free target window. It is computed with respect to a single defect cluster, without accounting for the other defects. A strategy that minimizes the local distance loss for a given cluster is called ``locally optimal." In \Cref{subsection:boundary_deformation}, we discussed how the strategies for each defect cluster are affected by the target window's boundary (see \Cref{fig:boundary_deformation}). A missing puzzle piece is the interdependence of the defect clusters when we evaluate the global code distance, accounting for all defect clusters and the deformation of the boundary. 
        
        Finding a combination of local strategies that maximizes the global distance is more complex than independently minimizing the local distance loss at each cluster. Local strategies at two different clusters can sometimes have cumulative effects on the code distance, e.g., if there is a minimal logical string passing through both clusters (see \Cref{fig:defective_grid}). This interaction between local strategies is complex and depends on the relative positions of all clusters. Therefore, to find the best global strategy, we need to consider multiple local strategies at each cluster. 
    
        Using this observation we summarize the steps of our algorithm in~\Cref{algo:snl}. First we identify the defect clusters in the extended window. We then make a list of local strategies for each defect cluster as in~\Cref{sec:neighboring_defects}. We define a global strategy by composing the local strategies and optimize boundary deformation for each as in~\Cref{subsection:boundary_deformation}. We return the global strategy resulting in the a surface code patch with the best distance (see \Cref{appendix:distance_calculation} for how it is computed). For simplicity we omit the heuristics (\Cref{subsec:heuristics}) from the pseudocode.
    
        \begin{algorithm}[ht!]
            \begin{algorithmic}
                \Procedure{SnL}{target window}
                \State
                \State $[\textit{D}_i] \gets \texttt{defect clusters in the extended window}$
                \State \texttt{FOR each defect cluster }$\textit{D}_i$:
                \State \hspace{1em} $[R_j] \gets \texttt{combination of ancilla repurposings}$
                \State \hspace{1em} $s_i \gets[\;]$ //\texttt{local strategies for} $\textit{D}_i$
                \State \hspace{1em} \texttt{FOR each 
     combination }$\textit{R}_j$:
                \State \hspace{2em} $s \gets \texttt{initialize strategy with } R_j$
                \State \hspace{2em} \texttt{Disable data qubits in invalid checks}
                \State \hspace{2em} \texttt{Build gauge checks} 
                \State \hspace{2em} $\texttt{IF valid stabilizers THEN add }  s \texttt{ to } s_i$
                \State
                \State $[\textit{S}_k] \gets \texttt{Cartesian product } \textit{s}_1 \times \textit{s}_2 \times \cdots$
                \State \texttt{FOR each global strategy }$\textit{S}_k$:
                \State \hspace{1em} \texttt{Remove checks outside the target window}
                \State \hspace{1em} \texttt{Move corners as needed}
                \State \hspace{1em} \texttt{Adapt remaining checks}
                \State \hspace{1em} \texttt{IF valid boundary THEN keep strategy}
                \State
                \State \texttt{Return global strategy with best distance}
                \EndProcedure
            \end{algorithmic}
            \caption{
            SnL achieves the optimal surface code distance in a target window of the defective processor.}\label{algo:snl}
        \end{algorithm}

        \subsection{Reducing the runtime and memory costs}

            Below, we discuss two key steps that improve the scalability of SnL.

            \subsubsection{Defects clustering}
        
                A key step of our algorithm is to define the smallest possible defect clusters which ensure the strategies for different clusters do not have overlapping checks. The smaller the cluster, the fewer the ways of repurposing, thus facilitating the search for local optimal solutions for each cluster based on the code distance loss. The method we use to define the smallest defect clusters is detailed in \Cref{appendix:pre_processing}. Put simply, we use DQD (with the link defects mapped to ancilla defects instead) to determine which qubits to group together since the super-stabilizers obtained with SnL are always strictly contained inside those defined by DQD (see \Cref{fig:primitive_operations}).

            \subsubsection{Heuristics}\label{subsec:heuristics}
        
                SnL relies on trying two choices of ancilla repurposing for each ancilla and link defect (depicted in \Cref{fig:snl_snakes}). We also consider treating each link defect as a data defect and applying the data defect primitive. We therefore find up to $2^a3^l$ unique strategies for a cluster with $a$ ancilla defects and $l$ link defects. While enumerating all possible strategies is feasible for small defect clusters, the runtime and memory overhead become prohibitive for large defect clusters. To reduce the computational cost of SnL, we choose to consider only a subset of the possible combinations of repurposings in each cluster. 
                
                We first sort all combinations of repurposing strategies so that the most promising ones are considered first, i.e., those that minimize the number of disabled data qubits and use the distance-preserving repurposings (see~\Cref{app:reducing_computational_costs}). We then iterate through them to remove the ``snakes'' and evaluate the local distance loss. In this process we skip some combinations or stop early based on conditions defined by several parameters (see~\Cref{app: heuristics_parameters}), most of which are related to the distance loss. Finally, we filter the strategies we found such that only a subset will be used for the global strategies. For example, only the locally optimal strategies for each cluster could be combined into a global strategy. 
                
                Our heuristics allow us to consider fewer strategies while guaranteeing that we evaluate those that most likely locally minimize distance loss. Depending on the choice of parameters, we ultimately find that the runtime and memory of SnL can be made comparable to what is achieved with DQD.

\section{Numerical Simulations}\label{sec:Results}

    In this section, we demonstrate how SnL leads to a significant improvement at all defect rates with respect to both the realized code distance and logical performance for arbitrary defect configurations relative to DQD, which includes the primitive from Ref.~\cite{auger2017fault} along with the bandage-like super-stabilizers in Ref.~\cite{wei2024low} and our optimized boundary deformation (i.e., we consider the best known implementation of DQD without weight-1 checks). Then, we also present numerical simulations for the logical performance, for the yield of preserving full distance, and for the logical CNOT gate based on lattice surgery.

    The setting we consider for the numerical results in \Cref{sec:code_distance,sec:logical_performance,sec:yield} is a target window as in \Cref{fig:sprinked_defects}(a) with dimensions $d_\text{targ} \times d_\text{targ}$, where $d_\text{targ}$ is the target code distance. We sample random configurations of defects and apply SnL or DQD to construct optimized code patches within the target window to carry out memory experiments.

    \subsection{Effective code distance} \label{sec:code_distance}
    
        In this section, we report the effective code distances achieved with SnL for random defect configurations and compare them to those obtained with DQD. In what follows, we denote the code distance of an adapted surface code by $d_\text{out}$ (output distance), the relative code distance by $\widetilde{d} = d_\text{out}/d_\text{targ}$, and their mean values over multiple samples by $\langle d_\text{out}\rangle$ and $\langle \widetilde{d}\rangle$. Unless otherwise specified, the value of $d_\text{out}$ we present will be the minimum of $d_X$ and $d_Z$.
        
        We randomly sample $1000$ defect configurations for each combination of defect rate ($q=0.001$ and $q=0.01$) and target distance $d_\text{targ}$ (using a $d_\text{targ} \times d_\text{targ}$ target window) ranging from $3$ to $69$, and generate adapted patches for each of them. Each qubit and link on the processor, including the padding ancillas, is independently assigned defective at the same probability, $q$, unless otherwise specified. We do not require the configurations to be unique.

        For distances up to $17$, we run SnL using a hierarchy of heuristics (culminating in the unconstrained global optimization) and report their performance in \Cref{app:heuristics}. We run the different heuristics in order of increasing runtime and stop for a given configuration when the runtime or memory usage become prohibitively large. We are able to run this on most defect configurations with 1\% defect rate up to distance $d=13$, but for larger distances we ran most configurations only with heuristics. We keep the code patch that has the largest $d_\text{out}$, and break ties using the sum of $d_X$ and $d_Z$, followed by the number of active data qubits. Moreover, with ancilla padding, we also have the option to use either half of the boundary ancillas for the native weight-2 checks in our code. Keeping the orientation of the logical operators fixed, this amounts to the freedom of reversing the Pauli type of checks. In order to produce optimal code patches, we consider both choices and record the best code distances.
        To save computational resources, we use stronger heuristics and do not consider both choices of boundary ancillas for distances starting from $21$ (see \Cref{app:heuristics} for details on the heuristics for large distances).

\begin{figure}[ht!]
            \centering
            {\includegraphics[width=.9\linewidth]{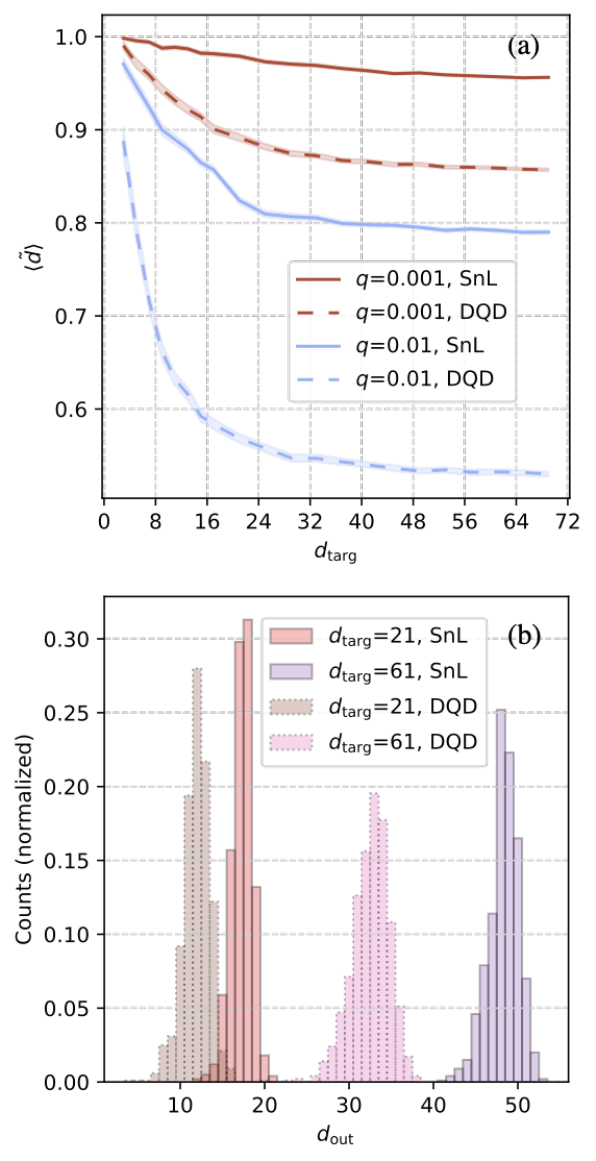}}
            \caption{(a) Average realized distance relative to the target distance, $\langle \widetilde{d}\rangle = \langle d_\text{out} \rangle /d_\text{targ}$. The maximum value is $1$ and indicates that we recover the full distance.
            (b) Histogram of realized distances for large distances and defect rate $q=0.01$. For SnL $\langle d\rangle$ is much greater and the distribution width is smaller compared to DQD.}
            \label{fig:relative_and_histogram}
        \end{figure}

        We report the average relative distance $\langle \widetilde{d}\rangle$ for different $d_\text{targ}$ in \Cref{fig:relative_and_histogram}(a). Each data point represents $1000$ random defect samples, with shaded regions indicating 95\% confidence intervals. We find that SnL significantly outperforms DQD, especially for higher defect rates. The relative distance trends to a constant as the target distance increases, indicating a stable performance trend at larger scales. For $q=0.01$, the relative distance by SnL converges to $0.79$ whereas DQD to $0.54$. The drop between the $q=0.001$ and $q=0.01$ curves for SnL is primarily due to data defects (see \Cref{app: code_distance} for details). We also plot the histogram of output distance for $d_\text{targ} = 21$, $61$ and for defect rate $q = 0.01$ in panel (b). The average output distance for SnL is clearly higher than for DQD, and importantly, the narrowing of the distribution indicates improved reliability of yield for patches with defects.

    \subsection{Logical performance}\label{sec:logical_performance}
    
        The effective code distance is a good indicator for the logical performance of the surface codes adapted with super-stabilizers. However, our protocol introduces a new primitive for defective ancilla qubits, so it is important to test that this primitive does not compromise logical performance. 
        
        Here we present Monte Carlo simulations of logical memory obtained with \texttt{stim}~\cite{gidney2021stim}. We use the \texttt{PyMatching 2} decoder~\cite{higgott2023sparse} and the SI1000 noise model~\cite{gidney2021fault}, which includes errors on idle qubits. Compared to the error models used by prior work, SI1000 is more realistic, especially because super-stabilizers introduce more idle locations on qubits. We run each simulation for $2d_\text{targ}$ rounds.

        We use a standard surface code syndrome extraction schedule (see \Cref{app:syndrome_extraction}) for all unmodified stabilizer checks. For the checks we modify as part of our defect strategy we retain the same schedule but without the gates that we remove. For example, around a data defect, the four neighboring ancilla qubits measure weight-3 gauge checks. The schedule for these weight-3 checks is the same as the ordinary schedule but with an idle round inserted where the fourth gate would ordinarily go. For repurposed checks we adopt the schedule of the defective ancilla. Therefore, from the point of view of the data qubits involved in the check, they are still measured in the same time steps they would ordinarily be measured. This way we prevent additional hook errors that might be introduced by modifying the check schedule.
        
\begin{figure}[t!]
            \centering
            \includegraphics[width=0.94\linewidth]{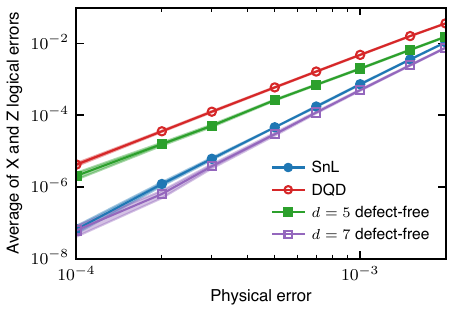}
            \caption{Logical performance in the sub-threshold regime. Different adaptations to the same distance-7 rotated surface code with a defective ancilla qubit.}
            \label{fig:ler_example_config}
        \end{figure}

        First, we test the distance-preserving repurposing in \Cref{fig:primitive_operations}(c) on a distance-7 rotated surface code that has a single defect on an ancilla qubit in the bulk. SnL retains the code distance of the patch, while DQD reduces the code distance to 5. As \Cref{fig:ler_example_config} shows, our adapted patch nearly matches the logical performance of the distance-7 defect-free code, and significantly outperforms both the distance-5 defect-free code and DQD. We note that DQD includes a pair of big super-stabilizers, so its logical performance could probably be improved by using the shells~\cite{strikis2023quantum} which we did not implement. However, even with the shells, the performance of DQD, especially at low physical error rates, would still be limited by its effective code distance. We expect it to remain similar to the logical performance of the distance-5 defect-free patch.

\begin{figure}[t!]
            \centering
            \includegraphics[width=0.98\linewidth]{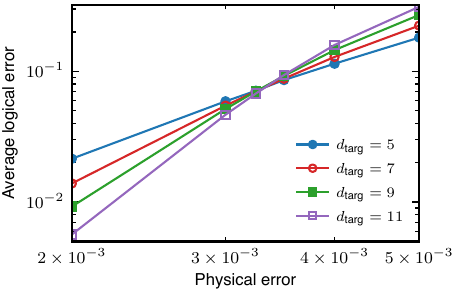}
            \caption{Threshold plot for the surface code adapted via SnL. The simulations are run for $2d_\text{targ}$ rounds with the SI1000 noise model. Each point is the average logical error rate (between $X$ and $Z$) on a sample of 1000 patches with $1\%$ defect rate.}
            \label{fig:threshold_plot_eg}
        \end{figure}

        We also study how defects influence the threshold of the code (see~\Cref{tab:thresholds},~\Cref{fig:threshold_plot_eg} and~\Cref{app:logical_plots} for the plots used to extract those values). We run logical memory experiments on the patches adapted to the defect configurations that we sampled at a defect rate of $1\%$, and average the results. When SnL is used, a defect rate of $1\%$ decreases the threshold of surface code (under the SI1000 error model) from $0.42\%$ in the defect-free case to $0.33\%$. However, with DQD, the threshold drops further to $0.27\%$. We also run a set of simulations using a standard depolarizing noise model\footnote{With two-qubit depolarizing noise on two-qubit gates, one-qubit depolarizing noise on one-qubit gates, and $X$/$Z$ noise on reset and measurement, all with the same error rate $p$.} that does not include idling errors. Under this model, SnL also results in a higher threshold than DQD.\footnote{We exclude defect configurations where our software fails to produce adapted patches for DQD. These excluded cases constitute $<1\%$ of the sampled cases for $d_\text{targ}=5$ and are very rare for higher $d_\text{targ}$; they also arise in prior works.}
        
        \begin{table}[t!]
            \centering
            \begin{tabular}{c|c|c}
                & Standard & SI1000\\
                \hline
                Defect-free & $0.86\%$ & $0.42\%$ \\
                \hline
                SnL (our approach) & $0.72\%$& $0.33\%$ \\
                \hline
                DQD (baseline) &$0.57\%$  & $0.27\%$ \\
            \end{tabular}
            \caption{Threshold values under the standard and SI1000 noise models, for the defect-free patches and patches with 1\% defect rates adapted with either SnL or DQD.}
            \label{tab:thresholds}
        \end{table}

    \subsection{Yield for preserving full distance}\label{sec:yield}

        We now shift our focus slightly to consider yield, which we define as the probability of preserving the full distance of the code ($d_\text{out}=d_\text{targ}$) in the presence of defects. Here we not only compare SnL with DQD within a target window of a larger surface code processor, but also study the impact of different padding strategies on a small chip where the target window is the entire chip. Here we compare ``no padding", wherein we consider a chip with only the boundary ancillas necessary for a standard rotated surface code, ``ancilla padding", wherein we have extra ancillas around the perimeter of our code as we do in the target window shown in \Cref{fig:sprinked_defects} and ``$d+1$", wherein we fabricate a patch of size $d_\text{targ}+1$, with only the necessary boundary ancillas, to increase the probability of realizing a distance-$d_\text{targ}$ patch. Ancilla-padding is more defect-tolerant than no padding because we are able to use the extra ancillas for ancilla repurposing in certain cases, but it requires an additional $2(d_\text{targ} - 1 )$ ancilla qubits. The $(d+1)$ approach enables us to tolerate any one data qubits defect and even some configurations of multiple data defects. This requires $4d_\text{targ} + 2$ more qubits than the no-padding approach.

\begin{figure}[t!]
            \centering
            \includegraphics[width=0.9\linewidth]{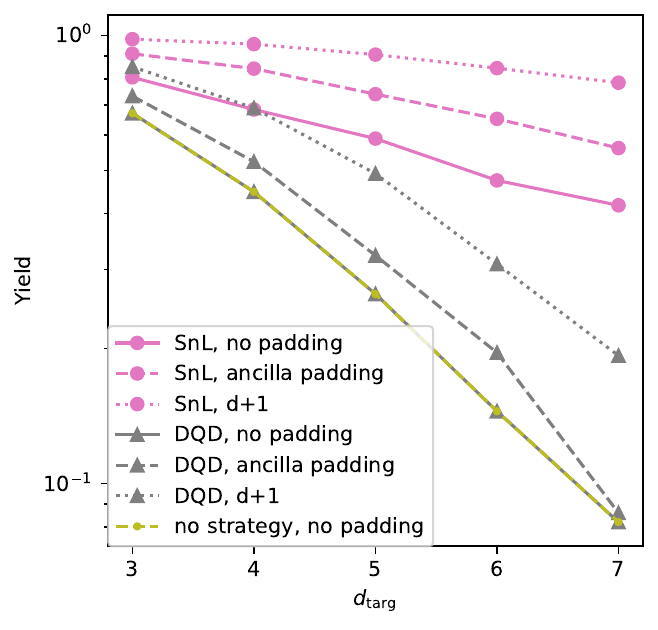}
            \caption{Yield of full-distance ($d = d_\text{targ}$) surface code for three different padding strategies: no padding, ancilla padding, and $d+1$. The defect rate is fixed at 1\% for qubit and link defects. ``no strategy, no padding" refers to the yield for a chip with no padding, where we discard the samples with any defective components.}
            \label{fig:yield}
        \end{figure}

        In \Cref{fig:yield}, we show the yields achieved by SnL and DQD with each padding strategy. For all padding strategies, SnL significantly increases the yield relative to DQD. With no padding, DQD obtains a 10\% yield at $d=7$, while SnL achieves a yield of 50\%. However, ancilla padding (which adds 12 ancilla qubits for $d=7$) increases the yield to 65\% for SnL. The ``$d+1$" padding strategy, which adds $30$ physical qubits compared to ``no padding'', further increases the yield to 80\%. Moreover, DQD with no padding has the same yield as using no strategy at all, because it cannot preserve the full distance in the presence of any defect. We observe that the padding strategies do offer the expected gains relative to the no-padding baseline. The gain for DQD with ancilla padding might seem surprising because DQD does not make use of ancilla repurposing. The reason for this gain is that with ancilla padding we have two choices for boundary ancillas. We choose one of two disjoint sets of alternating ancillas around the boundary. This gives us the flexibility to avoid boundary ancilla defects in some cases. For SnL we benefit from this effect as well as the additional ancilla repurposing possibilities.
    
    \subsection{Lattice surgery}\label{sec:cnot}
    
            \begin{figure*}[ht!]
                \centering
                \includegraphics[width=\linewidth]{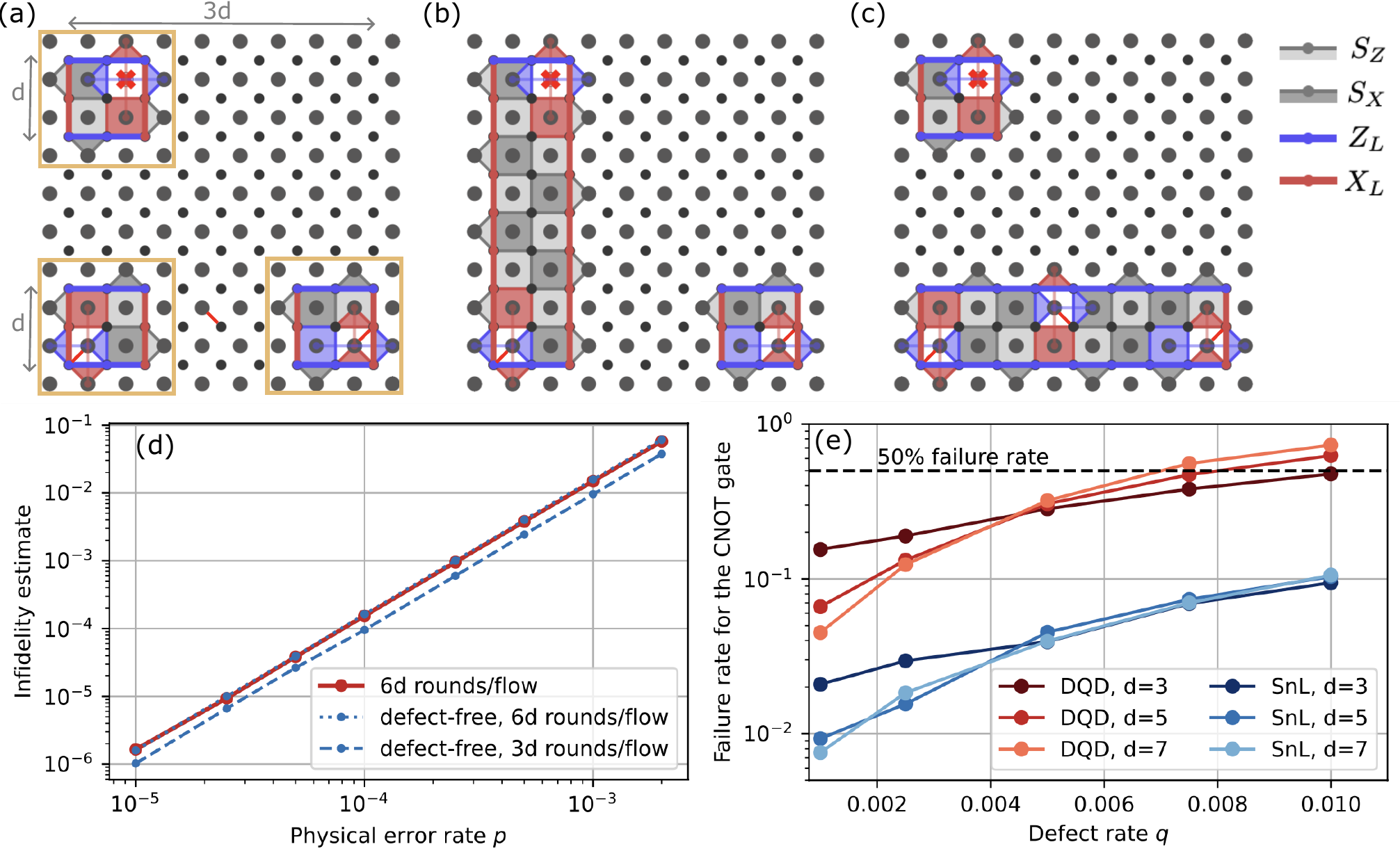}
                \caption{Lattice-surgery based CNOT gate with defects. (a) The control (top left), ancilla (bottom left), and target (bottom right) patches in a $3d \times 3d$ defective processor. The target window (in yellow) used for each patch has dimensions $d\times d$. Here $d=3$. (b) The merged control-ancilla patch (left). (c) The merged target-ancilla patch (bottom). (d) Estimate of the infidelity for the CNOT gate using the patches in (a). The errors are cumulated from the four flows ($XX$ and $ZZ$ merge/split experiments initialized and measured in the $X$ or $Z$ basis) over $6d$ (defective patches) or $3d$ (defect-free patches) measurement rounds per flow, where $d=3$. The CNOT gate itself takes place over $10d$ (defective case) or $5d$ (defect-free case) measurement rounds. (e) Yield experiment in a $3 d \times 3 d$ defective processor with the same layout as in (a) but where the control, ancilla and target patches are expanded into the routing space if needed to recover an effective distance $d$ for each patch. A failure occurs if any of the patches overlap.}           
                \label{fig:cnot}
            \end{figure*}

        Compared to DQD, SnL produces patches with higher distances in the presence of defects. Consequently, the logical operations between the patches have higher fidelity. Aside from this, SnL facilitates lattice surgery for two additional reasons. First, the boundaries of the adapted patch are less likely to be significantly deformed. Second, the holes in the routing space have smaller areas.
        
        In this section, we demonstrate that it is possible to realize a lattice-surgery-based logical CNOT gate between two surface code patches adapted using SnL. A basic example is shown in~\Cref{fig:cnot}(a)-(c). The three patches remain full-distance in the presence of ancilla and link defects, but it would not be possible to adapt these patches with DQD. We also show how to apply SnL to optimize for lattice surgery operations rather than individual logical qubits. In what follows, we now use $d$ to represent the target code distance for conciseness.
    
        First, we need to define the patches for the control, ancilla, and target qubits. To do so, we apply SnL to the entire processor, comprising the control, target and ancilla patches as well as the routing space, as if it was a single patch. Second, we choose a rectangular target window around each sub-patch, preserving the super-stabilizers we found for the entire processor, and re-compute the new optimal boundaries for that sub-patch. This guarantees that as long as the merged patch and the separate patches have minimum distance $d$, then the logical information is encoded with distance $d$ throughout the process. It is straightforward to define the boundaries of the sub-patches once the rectangular target windows containing them are defined, in order to optimize their effective code distance. After having defined the sub-patches for the control, target and ancilla qubits, shown in~\Cref{fig:cnot}~(a), we define the merged control-ancilla and ancilla-target sub-patches, illustrated in~\Cref{fig:cnot}~(b) and (c), respectively. We are now ready to set up the CNOT gate numerical simulation.

        We first stabilize the control, target and ancilla patches over $r=2d$ ($d$) measurement rounds, where $d$ is the code distance, for the defective (defect-free) patches. To realize a CNOT gate between the control and target patches, we then merge the control and ancilla patches \cite{horsman2012,Litinski2018latticesurgery}. We do so by measuring the product of their logical operators $Z_L Z_L$: we measure all Z-type stabilizers and super-stabilizers between the bottom $Z$ and top $Z$ logicals of the control and target patches, respectively. We then stabilize the merged control-ancilla patch over $2d$ ($d$) measurement rounds. We split the merged patch back into separate ancilla and control patches by measuring the data qubits in the routing space in the $X$ basis. We then stabilize the split patches over an additional $2d$ ($d$) rounds. We then merge the ancilla and target patches by measuring $X_L X_L$: this time, we measure all X-type stabilizers and super-stabilizers between the right $X$ and left $X$ logicals of the ancilla and target patches, respectively. We again stabilize the merged patch over $2d$ ($d$) measurement rounds. We now split the merged patch by measuring in the $Z$ basis.
        We then stabilize the codes one last time with $2d$ ($d$) measurement rounds. Overall, the gate takes $10d$ ($5d$) measurement rounds for the defective (defect-free) patches.
        
        Since the \texttt{stim} package that we use for numerical simulations does not allow nondeterministic observables, we split the CNOT experiment into four flows~\cite{gidney2023less}. We simulate lattice surgery between the control and ancilla patches, and between the ancilla and target patches, separately. We initialize and measure each pair of patches in either the $X$ or $Z$ basis. In each of the four flows, there are one or three deterministic observables. To estimate the infidelity of the logical CNOT as a function of the physical error, we sum up the average error rates for the observables from all flows and divide by two (averaging over the $X$ and $Z$ basis). See \Cref{app:cnot_infidelity} for more details. Each flow takes $6d$ ($3d$) rounds for the defective patches (defect-free patches): $2d$ ($d$) to stabilize the patches, $2d$ ($d$) during the merge and $2d$ ($d$) after the split.
                
        In~\Cref{fig:cnot}(d) we report the average error in the two lattice surgery experiments as a function of the physical error rate, for the defective patches shown in panels (a)-(c) and their defect-free counterparts, averaged over five million Monte Carlo samples. Each point is averaged over $1000$ random defect configurations. We find that the logical error rates are approximately the same (when rescaled to take into account the different number of error-correction rounds) for the defective and defect-free processors.
        For this simulation, we used the standard noise model.

        We now move to a yield analysis for the CNOT gate and compare SnL to DQD. Here we consider random defect configurations on a $3 d \times 3 d$ processor with different defect rates. Our goal is to realize a CNOT gate with three distance-$d$ surface code patches. We therefore define three $d\times d$ target windows at the top left, bottom left and bottom right corners of the processor such that we have the same layout as in \Cref{fig:cnot}(a). For each target window, we find the surface code patch with largest output distance $d_\text{out} = \min(d_X, d_Z)$, and expand the window by size $1$ vertically if $d_X<d$ and horizontally if $d_Z<d$. We repeat this process recursively until the algorithm yields three non-overlapping patches of effective distance $d_\text{out}\geq d$ and with sufficient room for routing (i.e., there must be at least one layer of data qubits separating each pair of patches). If we fail to produce such patches within the processor, we record a failure. We present the failure rate in \Cref{fig:cnot}(d) as a function of the defect rate, for different distances $d$. This numerical experiment is further detailed in \Cref{app:cnot_yield}. There are two crucial observations. First, at a 1\% defect rate, DQD results in a failure rate around 50\% for $d=3$, which grows as we increase $d$. Second, the failure rate for SnL does not grow with $d$, but remains below 10\%. Both trends can be explained by the relative distances achieved by the two methods as $d$ grows (\Cref{fig:relative_and_histogram}(a)).

\section{Conclusion}\label{sec:conclusion}

    In this article, we introduced SnL, a suite of novel and highly performant strategies for adapting the surface code to defective qubits and two-qubit gates. SnL relies on repurposing ancillas (for ancilla and link defects) to implement weight-2 gauge checks, allowing for reconstruction of missing stabilizers and helping preserve the distance of the adapted surface code. Additionally, all stabilizers and superstabilizers are measured over two error-correction rounds, just like in the state-of-art methods \cite{auger2017fault,strikis2023quantum,siegel2023adaptive,lin2024codesign,yin2024flexiscd,wei2024low}.
    
    Using SnL, we demonstrated a large gain in the distance of the adapted surface code over existing methods for code sizes up to thousands of qubits. For realistic defect rates the adapted surface code exhibits logical error rates close to the error rates of the defect-free surface code and a few orders of magnitude better than with DQD. SnL also greatly increases the probability of realizing a full-distance surface code. At the same time, computer resources needed for SnL can be on par with resources used in previous methods.

    To better appreciate the impact of SnL we also benchmarked the performance of a logical CNOT gate implemented via lattice surgery. We found similar performances for the adapted and defect-free surface codes with the same code distances. We also performed a yield estimation and found that SnL has a significantly smaller failure rate compared to DQD. This failure rate grows quickly with the distance for DQD whereas it remains approximately constant for SnL. Our results convincingly demonstrate that in the presence of defects  SnL shows a major improvement not only for memory storage, but also for logical operations.
    
    SnL enables more hardware-efficient quantum computation with the surface code when hardware components have a non-negligible probability of failure. Between variations in targeting, fabrication defects, and time-dependent TLSs, this will be the reality of superconducting qubit devices for the foreseeable future. Therefore, SnL can make fault-tolerant quantum information processing significantly more attainable and resource-efficient. For transparency and reproducibility, the code used for the numerical simulations is readily available on Github~\cite{SnL2024}.

    We identify a few directions of future work. First, in our logical error rate simulations, we measured $X$ and $Z$ gauge checks in alternating rounds. Measuring the same type of gauge checks for multiple rounds before switching can lead to better logical performance for large super-stabilizers \cite{siegel2023adaptive}. Second, SnL yields larger-distance surface codes than strategies involving weight-one gauge checks used in \cite{yin2024flexiscd,wei2024low}, however using weight-one gauge checks in conjunction with ancilla repurposing can give better distance for some defect configurations (see \Cref{app:weight1_checks}). Third, we constrained our ancilla qubits to be repurposed at most once. In some cases, however, it is possible to achieve larger distance by allowing ancilla qubits to be repurposed twice, at the expense of being able to measure all stabilizers and superstabilizers within two quantum error-correction rounds. Fourth, SnL may be optimized for biased noise, for instance by considering Clifford deformations~\cite{BonillaAtaides2021,Xu2023,Dua2024} of the adapted surface code. Further study of the logical performance under realistic noise assumptions is necessary to better understand the trade-offs involved.

\section{Acknowledgments}

    We thank Prasahnt Sivarajah, Arne Grimsmo, Colm Ryan, Will Morong and Fernando Brand\~{a}o for fruitful discussions about the algorithm and the experimental implementation of our strategies. We also thank Simone Severini, Bill Vass, James Hamilton, Nafea Bshara, Peter DeSantis, and Andy Jassy at Amazon, for their involvement and support of the research activities at the AWS Center for Quantum Computing.
    
    \emph{Note added.---}While finalizing the publication release process, we became aware of an independent work by Debroy et al.~\cite{debroy2024luci}, which addresses the problem of adapting the surface code in the presence of defective components.

\bibliography{bib_defects.bib}

\newpage
\onecolumngrid
\appendix

\section{Weight-1 checks}\label{app:weight1_checks}

    \begin{figure}[ht!]
        \centering
        \includegraphics[width=0.75\linewidth]{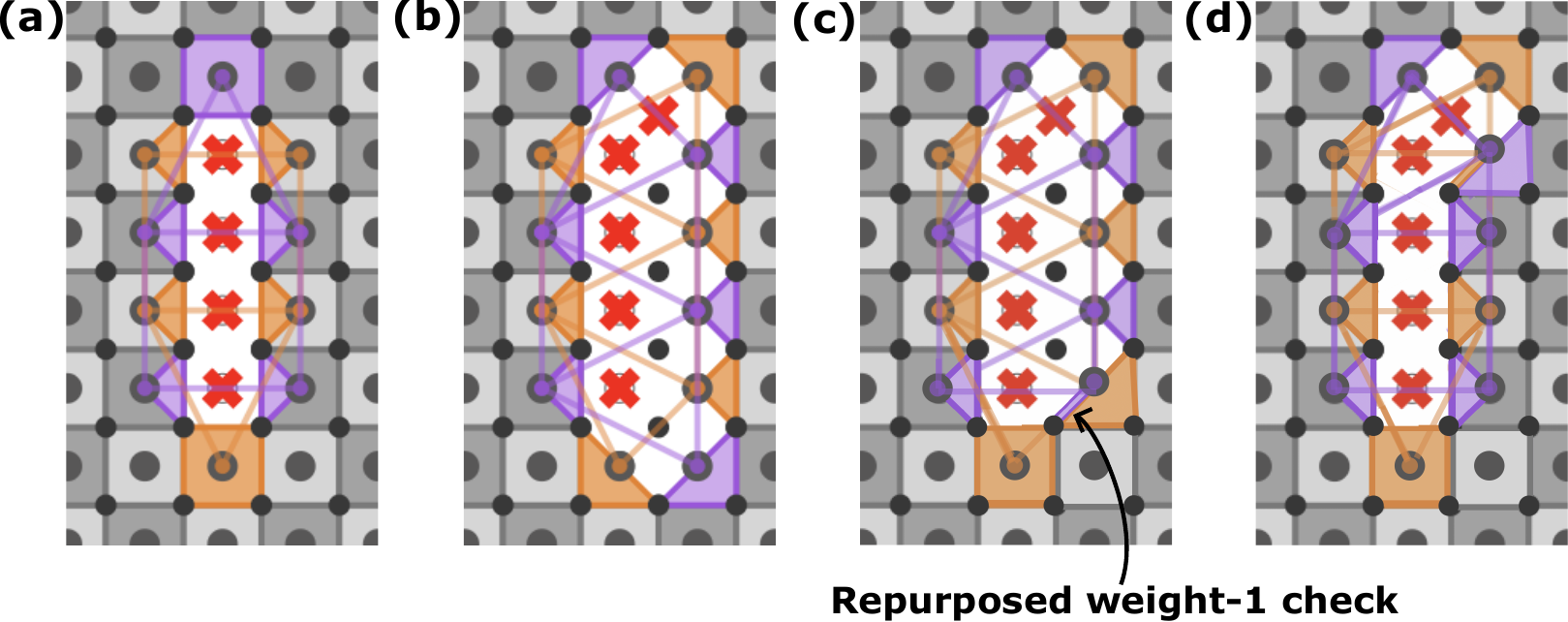}
        \caption{(a) Chain of adjacent ancilla defects. (b) Adding a data defect to the configuration in (a). (c) and (d) are examples where a repurposed weight-1 check could help the code distance vertically. However, (d) utilizes more checks overall. The weight-1 check is at the bottom right in (c) and top right in (d).}
        \label{fig:weight1_chain}
    \end{figure}
    
    Naively, weight-1 checks contradict the idea of having a stabilizer code by measuring a data qubit directly. However, Ref~\cite{wei2024low} suggests to enable weight-1 checks such as to prevent the ``avalanche'' effect that disables a large area of qubits. This avalanche effect results from the recursive search of weight-1 checks to disable, which ultimately leads to the holes to be rectangular in the rotated surface code. It follows that allowing weight-1 checks helps keeps more data qubits active. This was also studied in Ref~\cite{yin2024flexiscd}.
    
    The use of ancilla repurposing however already strongly suppresses this avalanche effect and in fact preserves significantly more data qubits. Indeed, with DQD this effect is usually triggered by ancilla defects since DQD disables all four neighboring data qubits around an ancilla defect. There are cases where weight-1 checks could be useful for SnL too, particularly when repurposing is not available, but it appears that they generally do not help the code distance and result in a larger error rate. Rather than allowing all gauge-1 checks, one should selectively allow them in cases where they are expected to result in gains, such as when all three data qubits around an ancilla are defective. We leave this addition for future work since the choice is not trivial.

    Here, we elaborate on why we decide not to allow repurposed weight-1 checks (i.e., when a repurposed weight-2 check is invalid because of a defective data qubit or a defective link). We however stress that this constraint could be relaxed given clearer rules on which repurposed weight-1 checks are beneficial for the code distance and which are not.
    
    When using SnL, long ``snakes'' of disabled data qubits arise when we have a chain of at least two adjacent ancilla defects such as in \Cref{fig:weight1_chain}(a). As long as there is a data defect on one side of the chain (see panel (b)), all the data qubits on that side will be disabled. On the other hand, allowing for repurposed weight-1 checks would allow us to disable only the data defect in that case (see panels (c) and (d)). If the data defect is at one of the endpoints of the chain, this increases the vertical code distance by 1, and otherwise by 2. This is one of the rare occasions where repurposed weight-1 checks help us recover more of the code distance. Now we notice that the code distance loss is the same for both panels (c) and (d), however, panel (d) has a lot more (repurposed) checks. Although they both have the same code distance, we would prefer (c) over (d) because the additional repurposed checks and the higher-weight stabilizer checks will introduce more noise to the stabilizer readouts. In fact, if more data defects are present along the chain, we would still prefer to use repurposed weight-1 checks only at the endpoints. This shows that in general, we do want to disable most data qubits along the ``snakes'' except potentially at the endpoints. 
    
    In summary, we think that disabling data qubits in ``snakes'' is generally a better strategy but emphasize that repurposed weight-1 checks could, in principle, improve the code distance in some cases.
    
\section{Ancilla repurposing} \label{app:efficient_check}

    \subsection{Repurposing on the boundary} \label{app:efficient_check_boundary}

\begin{figure}[ht!]
            \centering
            \includegraphics[width=.5\linewidth]{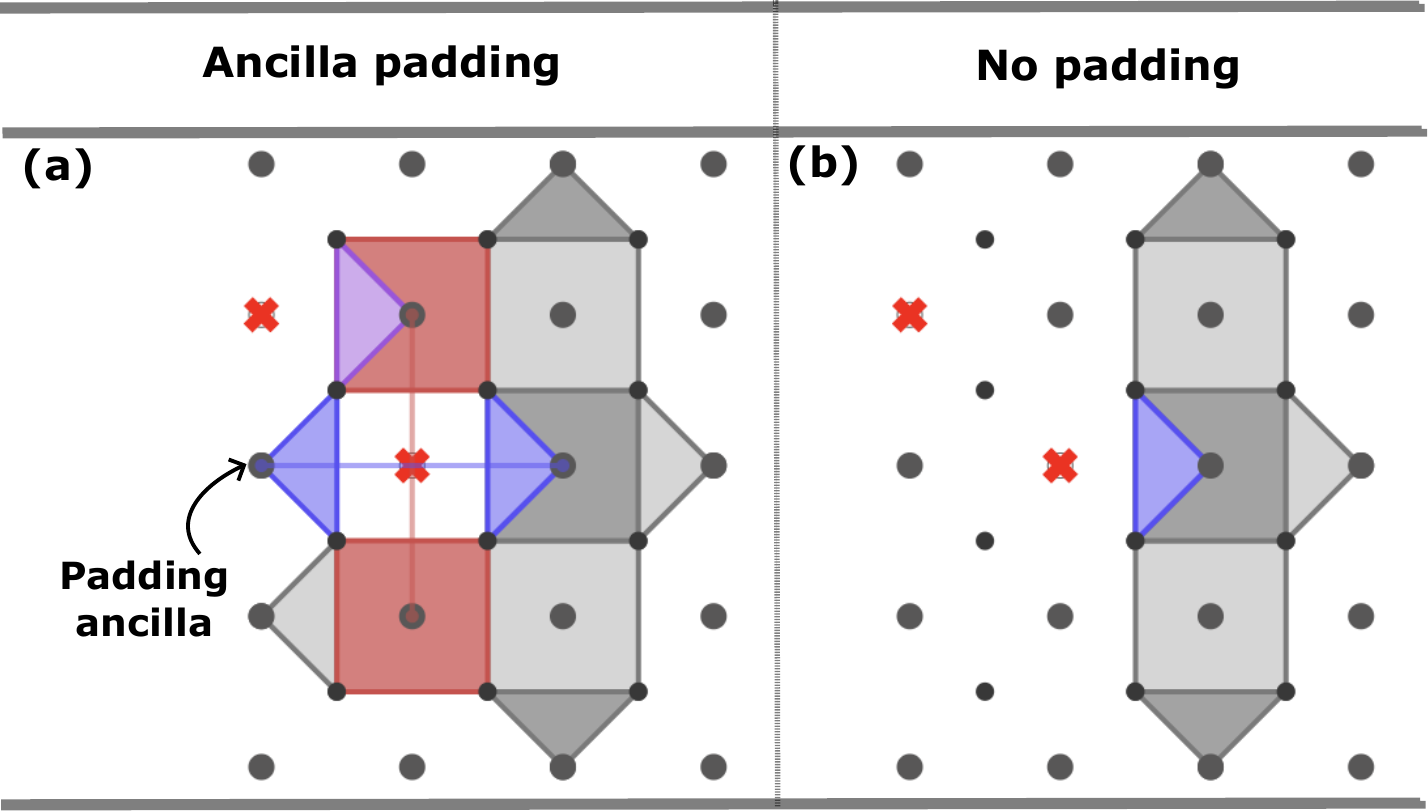}
            \caption{Ancilla repurposing for defective ancilla qubits on the boundary.}
            \label{fig:ancilla_padding}
        \end{figure}

        Here we show how ancilla repurposing applies to the defects on the boundary of the code patch. A defective ancilla qubit that measures a weight-2 stabilizer can be handled by repurposing its neighbor ancilla in the bulk to measure the check in alternate rounds. ~\Cref{fig:ancilla_padding}(a) illustrates this with a defect on the top left. If instead we have a defective ancilla qubit (such as the other ancilla defect in~\Cref{fig:ancilla_padding}(a)), we can use the ``padding'' ancilla qubit next to it for one of the repurposed weight-2 checks. Such an ancilla qubit is not part of the original surface code patch, but if it is present in our grid of qubits, we can use it to recover larger distance. The patch in ~\Cref{fig:ancilla_padding}(a) includes two defective ancilla qubits but suffers no distance loss. However, if the padding ancilla qubit was unavailable, the boundary deformation would lead to a distance loss as illustrated in panel (b).
    
    \subsection{Impact on the code distance} \label{app:efficient_check_distance}

\begin{figure}[ht!]
            \centering
            \includegraphics[width=0.5\linewidth]{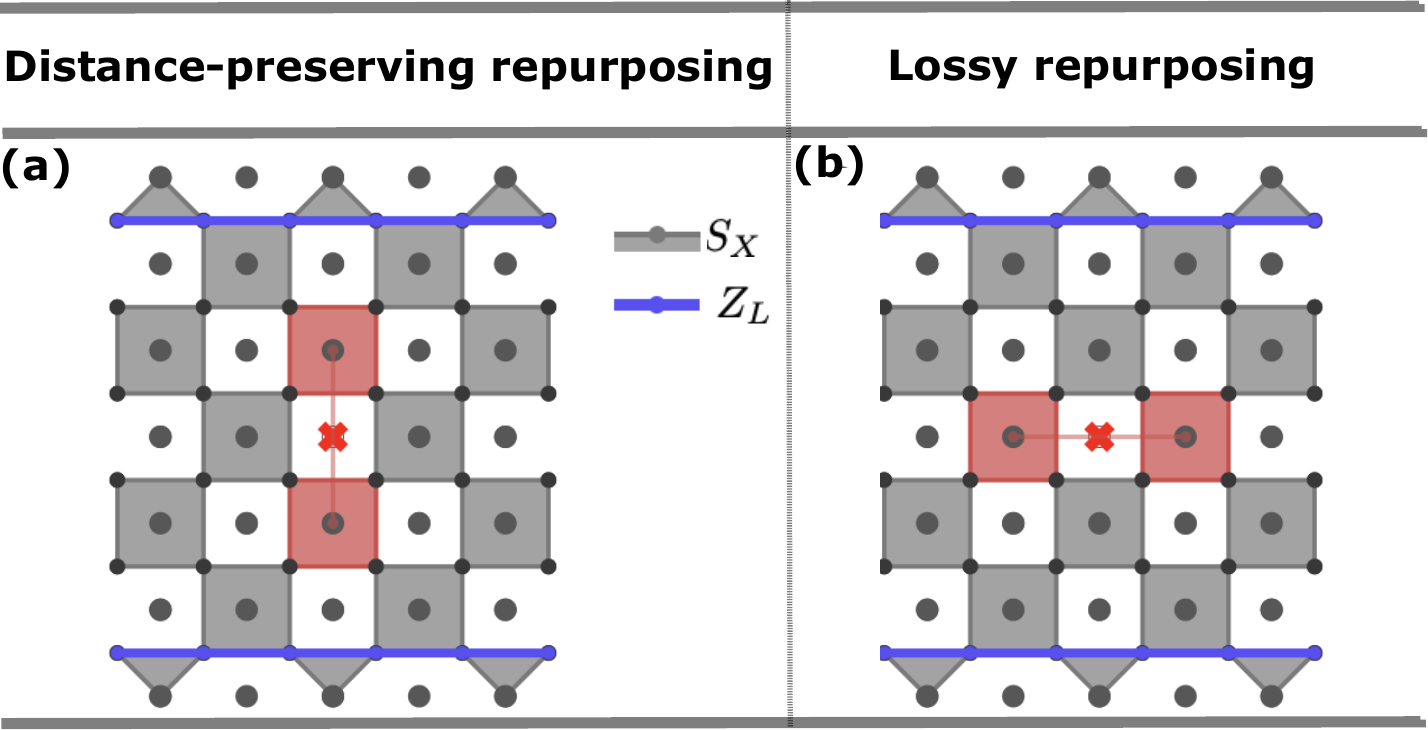}
            \caption{Distance drops when using repurposed checks for an ancilla defect for both orientations. One preserves the distance whereas the other leads to a distance drop by two in the direction along the weight-8 stabilizer.}
            \label{fig:distance_drop}
        \end{figure}

        We illustrate the impact of repurposed checks around an ancilla defect on the distance in~\Cref{fig:distance_drop}. The orientation for which the weight-8 stabilizer is aligned with the logical of opposite Pauli type results in a distance drop by 2 along that direction.
        
\section{Defining the super-stabilizers}\label{app:super-stabilizers}
    
    We define the super-stabilizers of $X$ (or $Z$) type by building a graph whose vertices are ancilla and data qubits. We add weightless edges between every defective/disabled data qubit and its neighboring ancilla qubits that would be used to measure $X$ (or $Z$) type stabilizers in the defect-free patch. For every repurposed check we also add an edge between the replaced ancilla and the repurposed ancilla if the replaced ancilla would measure a $X$ (or $Z$) type stabilizer in the defect-free patch, otherwise we add an edge between the pair of ancillas that is perpendicular to the pair of repurposed checks. Super-stabilizers are then defined by the connected components in this graph, i.e., we simply multiply the gauge checks and repurposed checks associated with ancilla contained in the connected component.

    We remark that we only keep strategies where the super-stabilizers satisfy all commutation relations. The graphs we build to obtain the super-stabilizers are also closely related to those used to compute the code distance (see \Cref{appendix:distance_calculation}). When defining the super-stabilizers for a given strategy, we can therefore simultaneously compute the distance loss for the cluster and check the validity of the strategy in terms of the commutation relations.

\section{Boundary deformation}\label{app:boundary_deformation}

        \begin{figure}[hbt!]
            \centering
            \includegraphics[width=\linewidth]{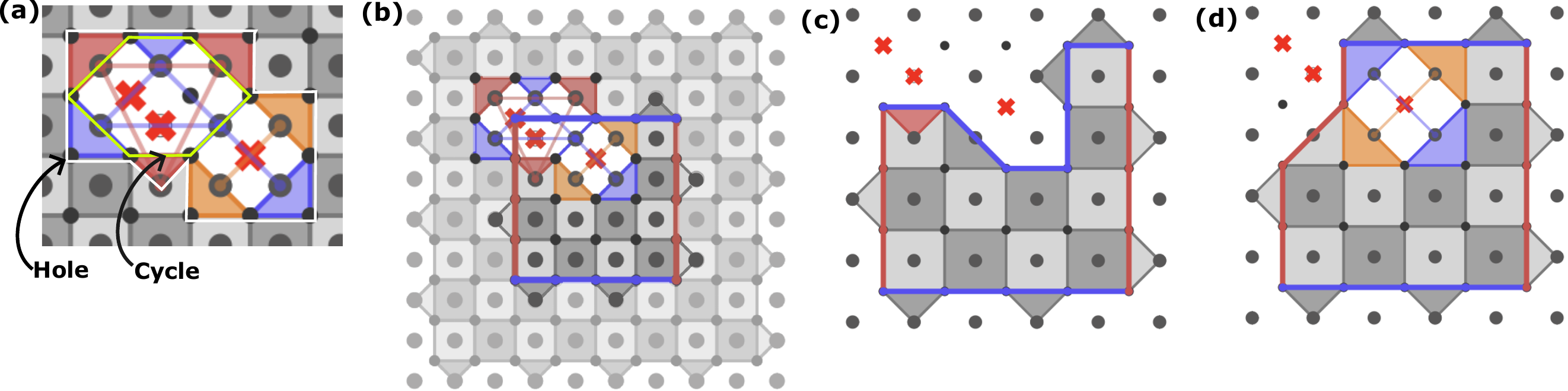}
            \caption{Definition of a hole and a cycle. (a) A hole (contoured in white) with two cycles. One of them is highlighted in green. (b) Hole enclosing the corner of the initial patch. The hole is \emph{open} because some of its gauges fall outside the target window, and removing them breaks the commutation relation between the super-stabilizers. (c)-(d) Two of the possible boundary deformations for the patch. (c) Gauges of the wrong type along both cycles are removed because they share a gauge of this type. (d) Only the gauges along the cycle enclosing the corner are removed.}
            \label{fig:holes}
        \end{figure}
            
        To facilitate the boundary deformation, we introduce the term \textit{hole} (see \Cref{fig:holes}(a)) to denote a set of inter-dependent super-stabilizers whose gauges commute with the checks outside the set but not with all the gauge checks within the set. Usually, a hole has a pair of $X$ and $Z$ super-stabilizers, but it can also have more super-stabilizers in complex cases. 
        
        We identify the \textit{holes} by grouping the super-stabilizers we found in the cluster based on commutation relations. We then need to determine which ones require boundary deformation because they contain gauge checks that are outside the target window. Such holes are denoted \textit{open holes}.
                
        Next we define the term \textit{cycle} (see \Cref{fig:holes}(a)) which is a closed and chordless contour with data qubits as vertices, that encloses only defective components. The gauge checks along each cycle are of alternating Pauli types and overlap at only a single data qubit. We clean the gauges around an open hole following one of the two possible cases. 

    \subsection{Edge hole}
            
        The first case is when the open hole does not involve a corner qubit, in which case we call it an edge hole. This is the simpler case and always leads to a single solution for boundary deformation. 
        
        First we identify the \emph{open} cycles. Similar to open holes, the open cycles are those with some gauge checks outside the target window. Along those cycles, we remove the gauge checks of the type opposite to that of the boundary. We then examine the remaining cycles in the hole to check if they had a gauge check removed; if so, they have become open and we also need to remove their gauge checks as well. 
        
        To simplify the process we therefore create a graph whose vertices are the cycles, and add an edge between every pair of vertices corresponding to cycles that share one (or more) gauge checks of the Pauli type to remove. All the connected components that include an open cycle (at the boundary) will tell us which cycles will become open. We now simply have to remove the gauge checks of the wrong type. The remaining gauge checks in those open cycles will become boundary checks in the deformed patch. 
        
        Finally, the gauges along the closed cycles will be combined into smaller super-stabilizers in the bulk.

    \subsection{Corner hole}
            
        The second and more complex case is when the open hole contains a corner qubit, in which case we call it a corner hole. We need to evaluate multiple possible boundary deformations, because we now need to define a new corner in the most optimal way possible. 
        
        We consider all the data qubits along open cycles and contained inside the target window as potential new corners. For each potential corner, we then remove all the gauge checks of the type opposite to that of the boundary. The approach is similar to the ``edge hole'' case.

        Our software does not handle the very rare case that the hole contains more than two corners. In such a case, the user can define a smaller target window to avoid the problematic defect cluster. 

\section{Computing the code distance}\label{appendix:distance_calculation}
    
    In order to calculate the code distances for both types of logical errors, we need to find the shortest horizontal and vertical logical operators. 
    
    Our method starts by building a graph whose vertices are the ancilla qubits used to measure super-stabilizers of the Pauli type $P$ opposite to the Pauli type $Q$ of the horizontal/vertical logical operators, and whose edges have weights equal to the Chebyshev distance between the nodes. We also add the virtual boundary nodes (which will be the endpoints of all logical strings in the graph sitting on the boundaries of the ideal defect-free patch) and the data qubits along the open holes on the two boundaries of Pauli type $Q$ to the graph (to compute the distance loss due to the boundary deformation). We then introduce a weight-0 edge between every pair of ancillas in every super-stabilizer as well as a weight-0 edge between every pair of repurposed and replaced ancillas in repurposed weight-2 checks. Finally, we add an edge between each virtual boundary node and every ancilla whose native stabilizer type is $P$ (i.e., we do not include repurposed ancillas) and every data qubit in the graph (due to boundary deformation) with its weight equal to the Chebyshev distance. The shortest logical length between the two virtual boundary nodes is then obtained using the Dijkstra algorithm.
    
    Further details can be found in our repository~\cite{SnL2024}. We use this approach both to 1) compute the code distance of the output patches and 2) compute the distance loss locally in a given defect cluster for a fixed combination of ancilla repurposings. The latter allows us to determine which local strategies are optimal when using the heuristics in \Cref{app:reducing_computational_costs}.

    Our method was validated with \texttt{stim} using the length of shortest graphlike error (i.e., using \texttt{shortest\_graphlike\_error()}) but requires less computing resources for large distances and high defect rates (resulting in a higher number of super-stabilizers) since we do not need to build a quantum circuit.

\section{Defect clustering}\label{appendix:pre_processing}

    First we virtually extend the defective processor (with two additional rows/columns of data qubits and ancilla qubits on each side) such that we treat boundary and bulk defects equally.
        
    Although at this step we do not distinguish between defects in the bulk and near the boundary, we do handle two types of defects specially. One is the defective padding ancillas that are on the processor but not part of the original code (see \Cref{fig:ancilla_padding}), and the other is the defective links that connect the original code to the padding ancillas. These components are not part of the original code. In both cases we ignore these defective components when defining the defect clusters in the extended patch, but will also avoid using these defective or disconnected padding ancillas for any repurposing strategy.

    Second we define the defect clusters. To guarantee that each hole is contained in one cluster, we consider the worst-case scenario where no repurposed check is implemented. This means that we consider only the DQD strategies in ~\Cref{fig:primitive_operations}. We build a graph where all the defective ancillas are connected to their neighboring data qubits, and all these data qubits as well as the defective data qubits are connected to their neighboring ancilla qubits. Finally, we connect all the ancilla qubits in the graph to their neighboring data qubits such as to reproduce the DQD checks in~\Cref{fig:primitive_operations}(a) and (b). The defect clusters are defined as connected components of this graph. 
    
    These clusters however often need to be recursively grown due to weight-1 checks (whose data qubits need to be disabled), an effect also known as the ``avalanche effect''\cite{wei2024low}. We remark that DQD holes are always rectangular in the rotated surface code since any concavity would result in weight-1 checks. This implies that each defect cluster can potentially cover a large area of the processor and we could ultimately find multiple small-size holes with our repurposing strategies within that area. Nonetheless, this is the only way to guarantee that strategies in individual clusters will never overlap.

    To limit the growth of the defect clusters, we remove nodes on the perimeter of the graph that we are certain will never be part of gauge checks because there isn't any first or second neighbor defect around those nodes. We refer the reader to our repository~\cite{SnL2024} for further details.
    
\section{Reducing the computational costs} \label{app:reducing_computational_costs}
            
    Here we discuss how to reduce the requirement of computational resources when we find optimal strategies for each defect cluster, without significantly lowering the quality of the algorithm's output patches. The desired balance between solution quality and search space size varies by problem, so we introduce multiple tunable parameters to enable a flexible trade-off.

    \subsection{Sorting all the combinations of repurposings} \label{app:sorting_repurposings}
    
        For a given defect cluster, the search for local strategies is implemented with a nested \texttt{for} loop. First we loop over each possible link assignment: whether we use ancilla repurposing or the data defect primitive for each link defect. Second, we loop over all possible combinations of repurposing strategies for ancilla and link defects. In order to avoid evaluating all these combinations, we want to break early from the nested loops. Then to avoid missing the locally optimal strategies, we need to sort the trial combinations so that the promising ones come first.
                
        First, we sort the link assignments that we iterate through in the outer loop. Since ancilla repurposing is usually the better approach for handling link defects than the data defect primitive, we start by applying ancilla repurposing on every link defect. After each iteration of the outer loop, we try the data defect primitive on one of the link defects and so on. Eventually, we try the data defect primitive on all link defects.
        
        Then we sort the combinations of repurposings for a fixed link assignment, which we go over in the inner loop. This is achieved by inspecting the two possible repurposings for each ancilla defect and link defect (that is not mapped to a disabled data qubit), and deciding which of them is more promising. Generally, we prefer the repurposing that is oriented to preserve the distance. However, not all repurposing checks survive after we remove the ``snakes'', so our decision also accounts for the likelihood that the repurposing checks survive. For each of the two possible repurposings around an ancilla or link defect, we check for defects on the qubits and links in the repurposing checks. This gives us an upper bound on the number of feasible weight-2 checks (more could be disabled later due to the chain effect). If one of the repurposings has fewer checks that are marked invalid, we mark it as the more promising one. If tied, we prefer the distance-preserving one.
                
        As a result of the sorting in both layers, the strategy that we first evaluates likely implements the highest number of repurposed checks and the most distance-preserving checks. The strategies that immediately follow implement a repurposing with flipped orientation for one ancilla or link defect, and far down the list are the strategies with some link defects adapted with the data defect primitive. This sorting guarantees that we encounter the optimal strategies for the defect cluster early in the nested loops. 
        
    \subsection{Parameters for the heuristics} \label{app: heuristics_parameters}
    
        Here we introduce what we call ``heuristics'', which are a set of conditions that allow us to either stop the search or skip some possibilities that are sub-optimal. We detail the different parameters for the heuristics that we use for reducing the computational costs. Some of them define stopping conditions for the nested \texttt{for} loop that we discussed in the previous section, and others define filters that further reduce the number of local strategies that we keep for each cluster. This rarely compromises the code distance due to the sorting we perform. 
        
        After executing the nested loops, we can filter out some local strategies before moving on to the global stage. This reduces the number of global strategies that we build from the local ones, hence reduces the memory and runtime of the algorithm at the cost of global optimization opportunities. This filter is defined by $\texttt{n\_sum}$, $\texttt{n\_max}$, and $\texttt{n\_sol\_max\_per\_cluster}$. The parameters $\texttt{n\_sum}$ and $\texttt{n\_max}$ set up a threshold for keeping a strategy, based on whether the sum (max) of the distance losses across the two dimensions is among the smallest $\texttt{n\_sum}$ ($\texttt{n\_max}$) different values from all strategies for the defect cluster. The other parameter, $\texttt{n\_sol\_max\_per\_cluster}$, directly caps the number of strategies that we keep for each defect cluster. When set to a small value, this parameter helps us dramatically reduce the computational costs. To balance the optimality and diversity of the local solutions we keep, we group the strategies by distance loss and select the ones with the most active qubits from each group. 
        
        When $\texttt{n\_sum}$ and/or $\texttt{n\_max}$ are used, we also apply them inside the nested loops, to determine whether to keep a strategy till the end of the loops.
        
        We define stopping conditions for the nested loops with the parameters $\texttt{link\_defect\_to\_ancilla}$, $\texttt{n\_zombie}$, and $\texttt{n\_skip}$. The first one, $\texttt{link\_defect\_to\_ancilla}$, is a flag that restricts the search for local strategies to those that strictly use ancilla repurposing for link defects (i.e., we do not use the data defect primitive for link defects). In this case we only iterate through the outer loop once, and do not consider mapping link defects to disabled data qubits. The second one, $\texttt{n\_zombie}$, provides a mechanism for breaking from the inner loop. It is triggered when we fail to find a better strategy (in terms of the number of functional data qubits) after evaluating $\texttt{n\_zombie}$ strategies for $\vec{\ell}$. The third one, $\texttt{n\_skip}$, defines another condition for breaking from the inner loop. It is triggered if the last $\texttt{n\_skip}$ items that we iterated through for the current $\vec{\ell}$ were all discarded due to not meeting the standard set up $\texttt{n\_sum}$ and/or $\texttt{n\_max}$.
        
        Finally, before applying boundary deformation to different combinations of local strategies, we can cap the number of these combinations by the parameter $\texttt{n\_sol\_max}$.
        
    \subsection{Limitations}
    
        It is more costly to search for optimal strategies for bigger defect clusters, since the number of possible solutions is exponential in the number of ancilla and link defects. After we apply the heuristics introduced above, we observe that the runtime and memory overhead of the algorithm still strongly depend on the size of the largest defect cluster in the patch.

\section{Heuristics used for sampling}\label{app:heuristics}

    Here we discuss the performance of different heuristics we tested when sampling target distances $3-17$ with a $1\%$ defect rate. We refer the reader to \Cref{app:reducing_computational_costs} for the different parameters used in the heuristics. The hierarchy of heuristics is shown in \Cref{tab:heuristics_hierarchy}. We ran the patches with $d = 21 - 69$ using the settings labeled 0, but modified to use $n_zombie= 2$.

    \begin{table}[ht!]
        \centering
        \begin{tabular}{c|c|c|c|c|c|c | c}
             Option / \# & 0 & 1 & 2 & 3 & 4 & 5 & 6\\
             \hline
             \emph{link\_defect\_to\_ancilla} & True & True & True & False & False &  False & False \\
             \hline
             \emph{n\_skip} & - & - & - & 10 & 10 & - & - \\
             \hline
             \emph{n\_zombie} & 1 & 10 & - & - & -  & - &- \\
             \hline
             \emph{n\_max} & 1 & 2 &  - & - & - & - & - \\
             \hline
             \emph{n\_sum} & 1 & 2 & - & - & - & - & - \\
             \hline
             \emph{n\_sol\_max\_per\_cluster} & 1 & 2 & 2 & 3 & 3 & 10 &- \\
             \hline
             \emph{n\_sol\_max} & 1 & 2 & 10 & 10 & - & 100 & - 
        \end{tabular}
        \caption{Hierarchy of heuristics that was used for the sampling of $d = 13 - 17$. The ``-" marks indicate that no value was set for this parameter, i.e., this condition was not used.
        }
        \label{tab:heuristics_hierarchy}
    \end{table}

    Heuristics $0 - 2$ ignore the data defect primitive for link defects, therefore reducing the number of strategies to consider per link defect from $3$ to $2$. Heuristic $0$ prevents any exponential blow-up in memory and runtime by allowing a single optimal strategy per cluster. Runtime is also limited by keeping only the first valid strategy found after storing all the possible combinations of repurposings. Heuristic $1$ allows for more exploration with $n_\texttt{zombie} = 10$ and keeps up to two strategies per defect cluster with a maximum of two unique max and sum distance loss values. We keep a maximum of two global strategies. This is leads to a smooth increase in memory and runtime. We now have a more global optimization of the code distance. Heuristic $2$ lifts all constraints except for the number of strategies per cluster, kept at two, but allow for $10$ global strategies. We allow for more optimization in terms of boundary deformation by having more global strategies to try. Heuristics $3-5$ allow for link defects to be mapped to data defects as well. Heuristics $3$ and $4$ but have a limited number of strategies per considered per link assignment ($n_\texttt{skip}=10)$ and allow for three strategies per cluster such as to still have global optimization. However, heuristic $3$ allow for $10$ global strategies whereas $4$ is unconstrained. Finally, heuristic $6$ is completely unconstrained, i.e., it looks for all possible global strategies. 

    \begin{figure}[ht!]
        \centering
        \includegraphics[width=\linewidth]{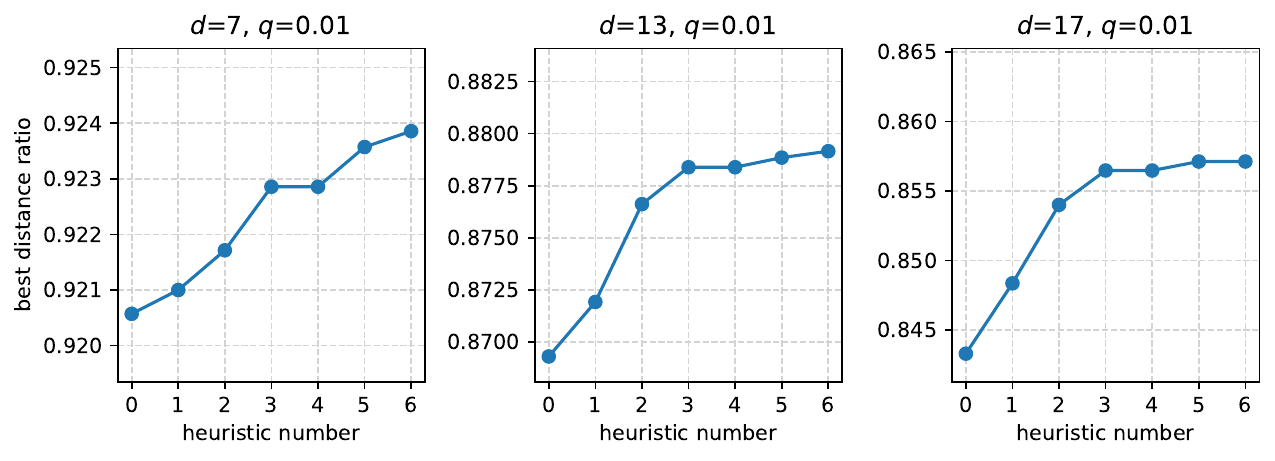}
        \caption{Average output distance divided by the target distance (here defined as the distance ratio) as a function of the heuristic number (defined in \Cref{tab:heuristics_hierarchy}) for target distance $7$, $13$ and $17$ at $1\%$ defect rate.}
        \label{fig:heuristics_convergence}
    \end{figure}

    When looking at the average output distance divided by the target distance (defined here as the distance ratio), and taking the best value from all prior heuristics and the current heuristic, we observe that the distance does increase smoothly with the heuristic number in \Cref{fig:heuristics_convergence} for target distances $7$, $13$ and $17$ at $1\%$ defect rate. In fact, more generally, heuristic $3$ is generally sufficient for target code distances above $11$. Indeed, a plateau can be observed for both $13$ and $17$. The main reason why this plateau is not present for distance $7$ is because boundary deformation is more impactful at smaller code distances. For that reason, it is useful to consider a greater number of global strategies. Nonetheless, we can see that even for distance $7$ the gain between heuristic $3$ and $6$ is about $0.1\%$. We also remark that the gain between heuristic $0$ and $6$ is on the order of only $0.3\%$ for distance 7 (because isolated defects are more common) but otherwise $1.0-1.5\%$ for distance $13$ and $17$. This implies that by doing only local optimization in the defect clusters we can generally recover most of the target distance. This is particularly true at small defect rates where defects are more likely to be isolated.
    
    \begin{figure}[ht!]
        \centering
        \includegraphics[width=.9\linewidth]{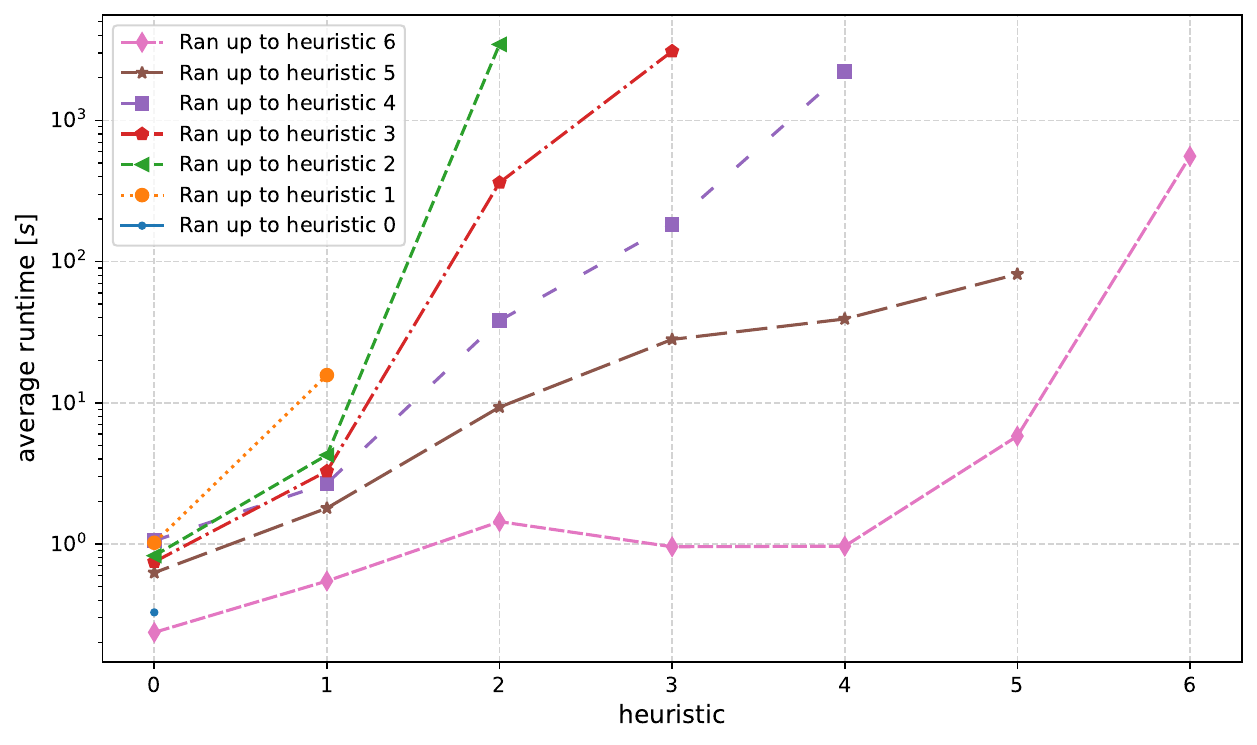}
        \caption{Average runtime in seconds per heuristic defined in \Cref{tab:heuristics_hierarchy}. The sample defect configurations are divided such that each curve has all the samples that ran up to the last heuristic that is plotted within $8$ hours. This is for target distances $3-17$ and $1\%$ defect rate.}
        \label{fig:heuristics_runtime}
    \end{figure}

    We now move to an analysis of the runtime. It is particularly challenging to have a good estimate for the runtime. This is because it strongly depends on 1) the number of defect clusters, 2) the average size of the defect clusters, 3) the number of ancilla and link defects, 3) the details of the sorting that we did in \Cref{app:sorting_repurposings}, 4) the optimization of corner placement during the boundary deformation. 
    A simple rule of thumb is that the larger the number of ancilla and link defects is per cluster, the more challenging the local optimization is and therefore, the longer the runtime. In \Cref{fig:heuristics_runtime} we plot the average runtime for distance $13$ as a function of the heuristic number. Here we divided the data into seven lines: the $n$th line has all the random defect samples that we could run up to heuristic $n$ under $8$ hours. First we remark that for all random samples, heuristic $0$ is fast (i.e., on the order of $1-10$s). Similarly, heuristic $1$ is generally guaranteed to run under $11$ min. Heuristic $2$ then shows a steep increase in runtime for some of the configurations. This is most likely a direct results of a lack of any stopping condition during the local optimization in each cluster. Heuristic $3$ now also considers mapping link defects to data defects which yields more strategies, however, it includes a stopping condition $n_\texttt{skip}$ for the local optimization (which explains the small drop in runtime for the longest line). However, for most of the lines, having more strategies per link defect increase the runtime, especially in presence of a significant number of link defects. Heuristic $4$ now allows for more global strategies that each require to run the boundary deformation part of the algorithm, which can become costly with corner holes. Generally, if heuristic $4$ can be run, heuristics $5$ and $6$ also can. We however decided to not run the next heuristic if the current heuristic took more than $20$ min to run.

    We remark that the standard variation is too high in \Cref{fig:heuristics_runtime} to be of any use or even shown. The main takeaway messages are that the larger the defect clusters, the more the ancilla and link defects, and the more corner holes the harder it is to run the algorithm without heuristics. It is most probable that the heuristics in \Cref{app:reducing_computational_costs} could be further refined to offer a better predictability for the runtime. It should however be noted that constraining the number of global strategies already strongly constrains memory usage.

    Furthermore, from \Cref{fig:heuristics_runtime,fig:relative_and_histogram}, it becomes clear that heuristic $0$ (the strongest heuristic that only uses local optimization) is already sufficient and results in a high plateau value of $0.79$ at large distances. Even at distances smaller than $11$, using no heuristic would only improve by $\sim 0.1$\% on average. For moderate distances we are talking about $\sim 1-2\%$ improvement. From the plateau in \Cref{fig:relative_and_histogram}(a), we can still expect only a few percents gain at larger distances. Finally, we also find that adding $n_\texttt{skip}$ to heuristic $1$ and $2$ can strongly help reduce the runtime with compromising the performance of the algorithm. We however leave a more detailed analysis of the heuristics proposed here for future work.

\section{Additional numerical results}\label{appendix:numerical_results}

    In this appendix, we report additional numerical results from sampling done in \cref{sec:code_distance,sec:logical_performance}. 

    \subsection{Code distance} \label{app: code_distance}

\begin{figure}[ht!]
            \centering
            \includegraphics[width=.7\linewidth]{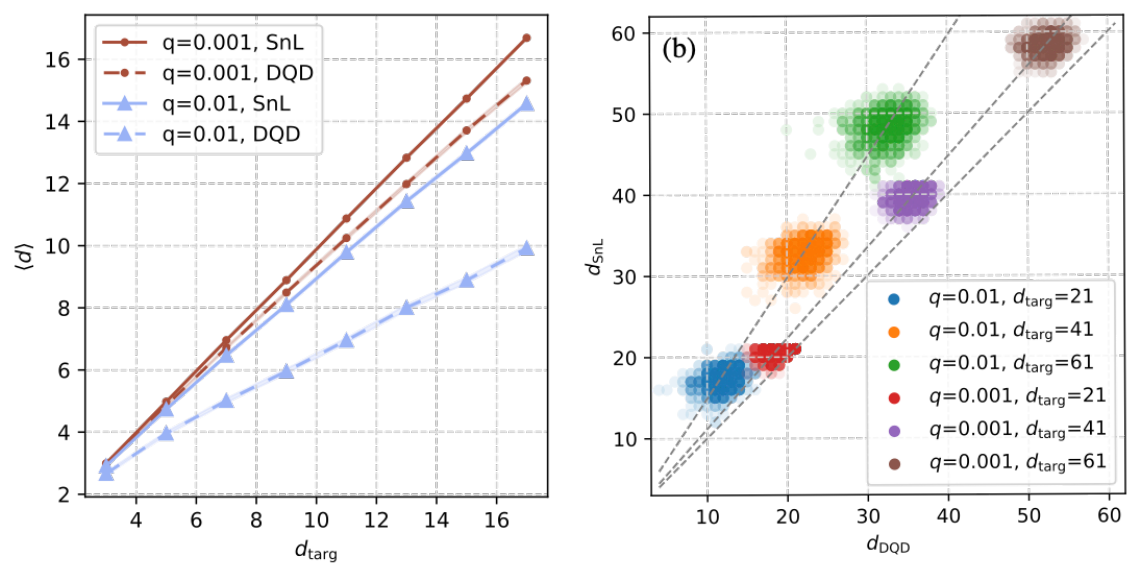}
            \caption{(a) Average $d_\text{out}$ for $d_\text{targ}$ ranging from 3 to 17.
            (b) Correlations between the code distances outputted by SnL and DQD. 
            We see that for all defect configurations SnL outperforms DQD. 
            The dashed lines have slopes of $1$, $1.12$, and $1.49$, where the latter two are based on the large $d_\text{targ}$ plateau values from~\Cref{fig:relative_and_histogram}.} 
            \label{fig:small_distance}
        \end{figure}

        Here we focus on the code distances that were obtained in \Cref{sec:code_distance}. In \Cref{fig:small_distance}(a) we report the average output distance $d_\text{out}$ as a function of the target distance $d_\text{targ}$ for both SnL and DQD, at $q=0.001$ and $q=0.01$ defect rates.

        In \Cref{fig:small_distance}(b), we plot the correlation between the distances outputted by SnL and DQD. For defect configurations sampled at different values of $q$ and ${d_\text{targ}}$, the SnL algorithm consistently outperforms DQD, and the ratio between the two outputted distances remain nearly constant when $q$ is fixed and $d_\text{targ}$ is varied.

\begin{figure}[ht!]
            \centering
            \includegraphics[width=.9\linewidth]{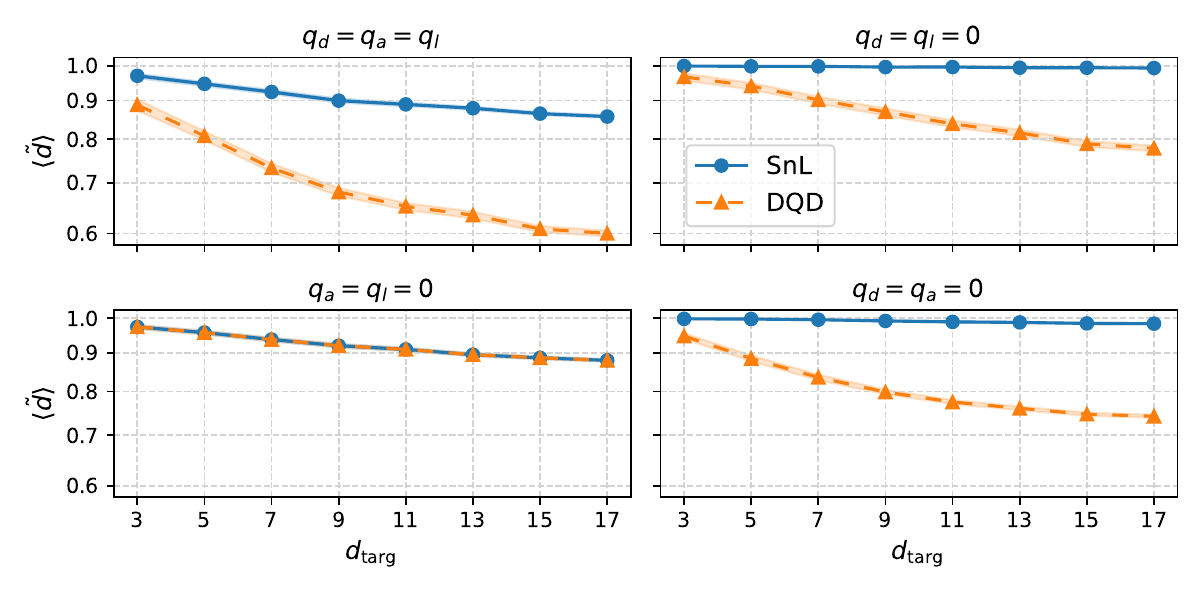}
            \caption{The average code distance for SnL and DQD for code distances from $3$ to $17$. Each plot shows a different ratio of defect probabilities for the three types of defects, data, ancilla, and link. In each case, the non-zero probabilities are set to 1\%. Notice that when we have only data defects, the two approaches are identical. SnL offers the greatest gains when we have link and ancilla defects in the code patch. Each average is composed of 1000 randomly sampled defect configurations.}
            \label{fig:differentratiosaverage}
        \end{figure}

        We have thus far focused on the case where the probabilities of the three types of defects, data, ancilla, and link are all equal. Let $q_d$, $q_a$, and $q_l$ be the data, ancilla, and link defect rates, respectively. We also studied the case where we have only one type of defect and the other two types of defects have zero probability, e.g. $d_q = 0.01, \, d_a = d_l = 0$. We sampled 1000 random defect configurations for each distance and applied both our SnL algorithm and DQD to each configuration. \Cref{fig:differentratiosaverage} shows the average distance relative to the designed distance with different ratios of defect rates. The non-zero defect rates are set to 1\%. With data, ancilla, and link defects at equal rate, we see that SnL outperforms DQD increasingly at larger distance. This reflects how DQD performs particularly badly for large defect clusters, which become more common as we increase the code patch size. With only data, link, or ancilla defects, we see greater relative distance for both strategies compared with $q_d = q_a = q_l = 0.01$. This is expected because we have few defects in total. Note that with only data defects, SnL and DQD are identical approaches. The distance loss in this case ($q_d = 0.01, \, q_a = q_l$) is close to that of SnL with $q_d = q_a = q_l = 0.01$. This demonstrates how the presence of ancilla and link defects has relatively little effect on the average distance with SnL. With only link or only ancilla defects, SnL recovers nearly the full distance, while DQD gives a significant drop in distance.

    \subsection{Logical performance}\label{app:logical_plots}

        We now move on to the analysis of the logical performance (see \Cref{tab:thresholds} for the threshold values). Here we consider two noise models: a standard depolarizing noise model and SI1000~\cite{gidney2021fault}.

        \subsubsection{Standard depolarizing noise model}

\begin{figure}[ht!]
                \centering
                \subfloat[SnL]{%
                \includegraphics[clip,width=0.49\linewidth]{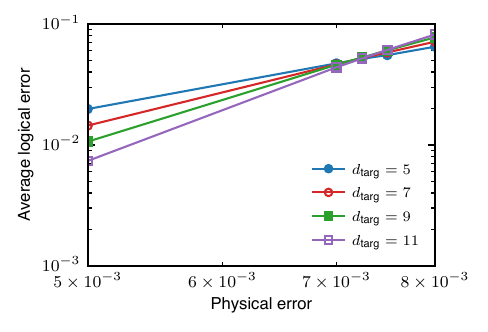}%
                        }
                        \subfloat[DQD]{%
                \includegraphics[clip,width=0.49\linewidth]{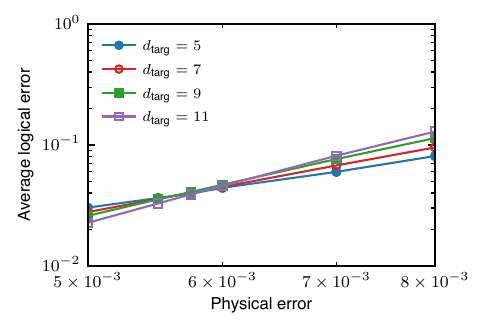}%
                        } \\
                        \subfloat[Defect-free]{%
                \includegraphics[clip,width=0.49\linewidth]{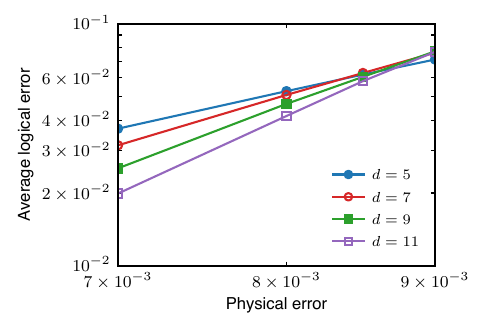}%
                        }
                \caption{Threshold plots with the standard noise model.}
                \label{fig:thresholds_standard}
            \end{figure}

            The threshold plots for SnL (our approach), DQD (baseline) and the defect-free case using the standard depolarizing noise model are shown in \Cref{fig:thresholds_standard}. The simulations are run for $2d_\text{targ}$ rounds. Each point is the average logical error rate (between $X$ and $Z$) on a sample of 1000 patches with $1\%$ defect rate.

        \subsubsection{SI1000 noise model}

\begin{figure}[ht!]
                \centering
                \subfloat[SnL]{%
                \includegraphics[clip,width=0.49\linewidth]{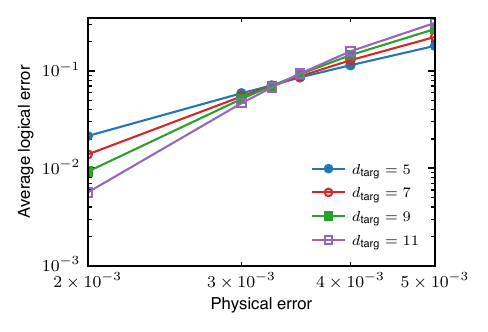}%
                        }
                        \subfloat[DQD]{%
                \includegraphics[clip,width=0.49\linewidth]{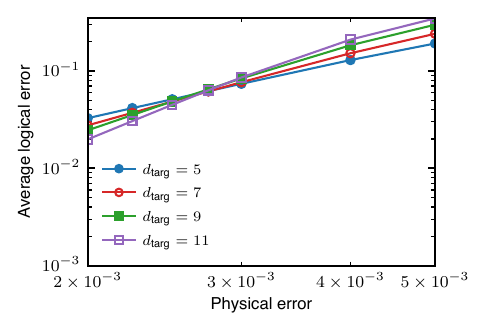}%
                        } \\
                        \subfloat[Defect-free]{%
                \includegraphics[clip,width=0.49\linewidth]{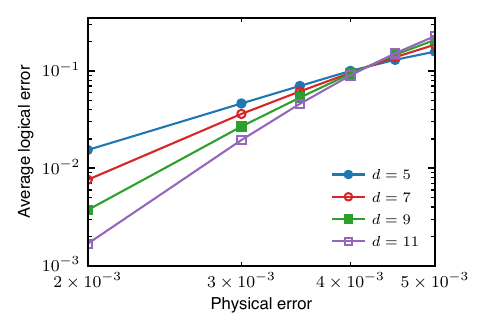}%
                        }
                \caption{Threshold plots with the SI1000 noise model.}
                \label{fig:thresholds_si1000}
            \end{figure}

            We use the same methodology as for the standard noise model but now consider the SI1000 noise model, a more realistic noise model for superconducting circuits. The resulting threshold plots are shown in \Cref{fig:thresholds_si1000}.

            An important difference between the SI1000 noise model and the standard noise model, beyond measurement and reset errors, is the presence of noise during idling. Requiring more measurement rounds to measure all super-stabilizers also lead to more idling errors.

\begin{figure}[ht!]
            \centering
            \includegraphics[clip,width=0.60\linewidth]{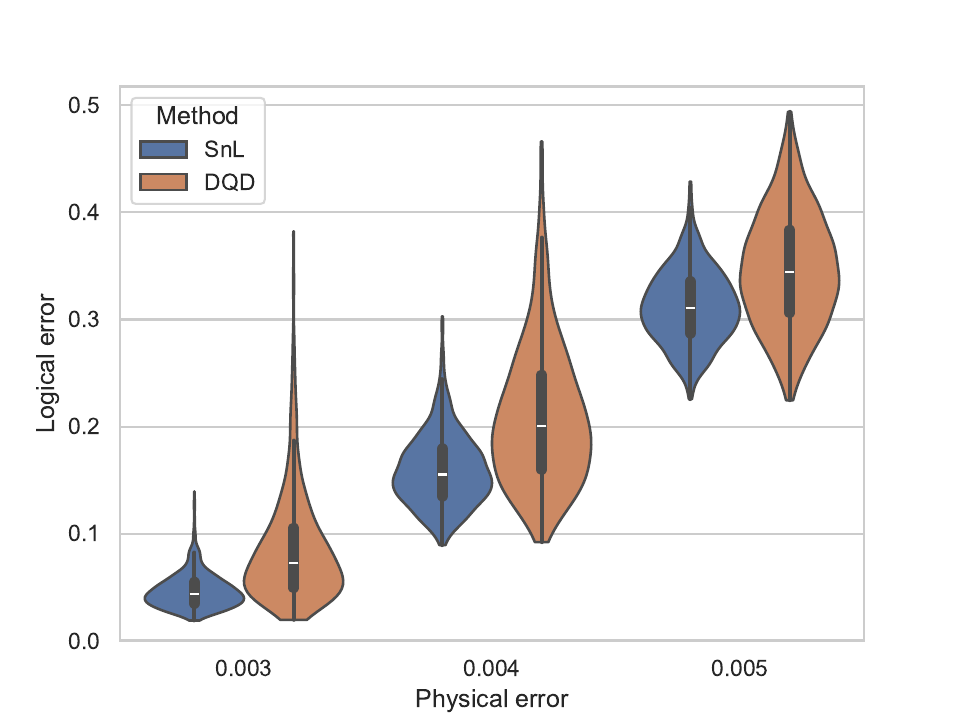}%
            \caption{Distribution of logical errors with the SI1000 noise model.}
            \label{fig:violin}
        \end{figure}

            Finally, we show the distribution of logical errors using a ``violin'' plot in \Cref{fig:violin}. In each ``violin'', the white dot is the median, the black box is the inter-quartile range (IQR), and the black line extends to $1.5\times$IQR. We observe a wider distribution of logical errors for DQD, relative to SnL, as expected from the width of the histograms of the output code distances in \Cref{fig:relative_and_histogram,fig:small_distance}.

            \newpage

\newpage
    
\section{CNOT gate}

    In this appendix, we explain the methodology behind \Cref{fig:cnot}.
    
    \subsection{Estimate of the infidelity} \label{app:cnot_infidelity}

        We now discuss how we estimate the CNOT gate infidelity shown in \Cref{fig:cnot}(d). To avoid non-deterministic observables in \texttt{stim}, we divide the lattice surgery CNOT into four flows in a manner similar to the approach in Ref.~\cite{gidney2023less}. Put simply, we consider the two merge-split measurements (control-ancilla and ancilla-target) separately, and for each of these we initialize and measure the patches both in the $X$ or $Z$ basis. To determine the errors in each flow, we measure the deterministic observables summarized in \Cref{tab:cnot_observables}. Taking the sum of the errors in all the deterministic observables in \Cref{tab:cnot_observables}, divided by two for the $X$ and $Z$ bases, approximately gives the infidelity of the CNOT gate.

        \begin{table}[ht!]
            \centering
            \begin{tabular}{c|c|c}
                Merge-split experiment & $Z$ init/meas basis & $X$ init/meas basis\\
                \hline
                Control-ancilla ($ZZ$) &  $Z_CZ_A$, $Z_C$, $Z_A$ & $X_{C \oplus A}$ \\
                Ancilla-target ($XX$) & $Z_{A\oplus T}$ & $X_AX_T$, $X_A$, $X_T$
            \end{tabular}
            \caption{Deterministic observables for the four flows. Here $X_i$ means the $X$ operator on the qubit $i$, where $i = C$ (control), $A$ (ancilla) or $T$ (target). Moreover, $X_{i \oplus j}$ refers to the $X$ operator of the patch resulting from merging patches $i$ and $j$. Alternatively, $X_{i \oplus j}$ corresponds to $X_i X_j$ after applying a $Z$ correction on $X_j$ depending on the measurement outcome of the $X$ string in the routing space during the split~\cite{horsman2012}.}
            \label{tab:cnot_observables}
        \end{table}

\begin{figure}[ht!]
            \centering
            \includegraphics[width=0.5\linewidth]{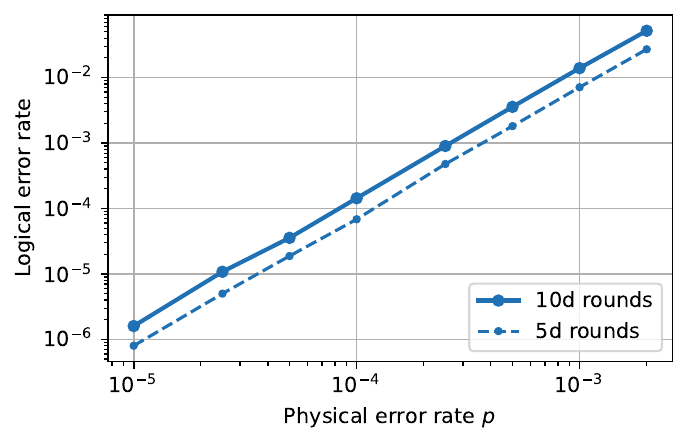}
            \caption{Logical error rate (in the shortest/horizontal direction) for the distance-$(3, 7)$ defect-free surface code over $10d$ and $5d$ measurement rounds, for $d = 3$, corresponding to the number of measurement rounds for the CNOT gate in \Cref{fig:cnot}(a)-(d).}
            \label{fig:cnot_3x7}
        \end{figure}

        We also expect this quantity to be close to the logical error rate for a memory experiment with a distance-$(3, 7)$ surface code over the same number of measurement rounds. This is because the performance of the merged patches mostly determines the infidelity of the CNOT gate in this case. This is confirmed in \Cref{fig:cnot_3x7}(d): the worst logical error rate for $10d$ and $5d$ measurement rounds (corresponding to the number of rounds for the defective and non-defective cases, respectively), with $d=3$, match with the estimate for the CNOT infidelity for both the defective and non-defective cases in \Cref{fig:cnot}.

    \subsection{Yield experiment} \label{app:cnot_yield}

        We now explain how we set up the yield experiment in \Cref{fig:cnot}(e). We consider a $3 d_\text{targ} \times 3 d_\text{targ}$ processor such that three patches with code distance $d_\text{targ}$ can fit side by side in both directions. We position the control, ancilla and target patches as in \Cref{fig:cnot}(a): the control patch is at the top left corner, the ancilla patch is at the bottom left corner, and the target patch is at the bottom right corner. The merged patches are defined on the left and bottom halves of the processor as shown in panels (b) and (c). 
        
        In this experiment, we ultimately want three distance $d_\text{targ}$ patches to operate so that they realize the CNOT gate. However, a random defect configuration
        generally results in distance loss. To recover the logical performance of the individual patches, we might want to extend them into the routing space such that their distance is $d_\text{targ}$. There is however a limit to how much we can grow the three patches into the available routing space. The goal of this experiment is to determine how often can we recover three distance $d_\text{targ}$ patches in a $3 d_\text{targ} \times 3 d_\text{targ}$ processor to realize an operation such as the CNOT gate. This experiment fails when the three resulting patches overlap (i.e., if there is not at least one row/column of data qubits separating the patches). The failure rate is shown in \Cref{fig:cnot}(d) as a function of the defect rate, for different target distances. 
        
        We remark that we always find the optimal strategy for the entire processor first, and then define subpatches with optimal boundary deformation, as it was done for \Cref{fig:cnot}(a)-(c). First we define a target window for the entire processor and find the strategy yielding the best code distance overall. We then define three rectangular target windows for the subpatches: $w_C$, $w_A$ and $w_T$ for the control, ancilla and target qubits, respectively. $w_C$/$w_A$/$w_T$ has its top left/bottom left/bottom right corner fixed at the top left/bottom left/bottom right corner of the processor's window. They each start with the minimum area such that they and contain ideal $d_\text{targ}$-surface code patch. We iteratively grow them in both directions (adding one row/column of data and ancilla qubits to the subpatch's target window at a time) until the surface code patch contained in the windows have minimum distance $d_\text{targ}$. We stop and declare a failure as soon as there is no longer a row/column of data qubits left between the patches since routing would then be impossible.

        Using DQD always leads to bigger distance loss, and consequently, requires digging more into the routing space. It is not surprising that the failure rate is significantly higher for DQD than for SnL. What is however interesting is that the failure rate grows quickly with the target distance for DQD. Intuitively, this is because more defective components are found in bigger codes and the holes end up being generally bigger. This implies that having a fixed $3 d_\text{targ} \times 3 d_\text{targ}$ processor is generally not sufficient to have a successful experiment. SnL, on the other hand, shows a near constant failure rate with increasing distance. This results from having smaller holes and distance-preserving strategies. This analysis could be extended to larger code distances with SnL and to different processor sizes. The bigger the processor size for fixed $d_\text{targ}$ is, the smaller the failure rate.

\section{Syndrome extraction schedule}\label{app:syndrome_extraction}

\begin{figure}[ht!]
        \centering
        \includegraphics[width=0.5\linewidth]{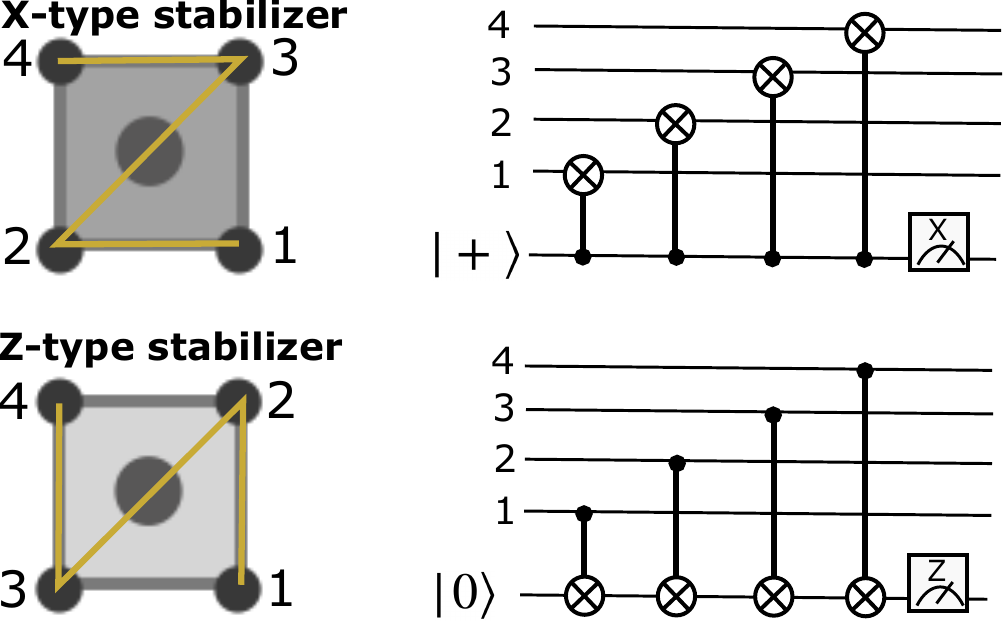}
        \caption{Syndrome extraction schedule for $X$ and $Z$ type stabilizers.}
        \label{fig:gates_schedule}
    \end{figure}

    The syndrome extraction extraction schedule we use for the numerical simulations is shown in \Cref{fig:gates_schedule}. We remark that flipping the $X$ and $Z$ boundaries of the rotated surface code in \Cref{fig:sprinked_defects}(b) also swaps the schedules of $X$ and $Z$ type stabilizers.
    
\end{document}